\documentclass[12pt]{article}
\usepackage{epsfig}
\usepackage{amsmath}
\usepackage{hhline}
\usepackage{amssymb}
\usepackage{times}
\usepackage{cite}
\usepackage[]{lineno}
\usepackage{rotating}

\newlength{\dinwidth}
\newlength{\dinmargin}
\begin{document}  
\newcommand{\pom}{{I\!\!P}}
\newcommand{\reg}{{I\!\!R}}
\newcommand{\slowpi}{\pi_{\mathit{slow}}}
\newcommand{\fiidiii}{F_2^{D(3)}}
\newcommand{\fiidiiiarg}{\fiidiii\,(\beta,\,Q^2,\,x)}
\newcommand{\n}{1.19\pm 0.06 (stat.) \pm0.07 (syst.)}
\newcommand{\nz}{1.30\pm 0.08 (stat.)^{+0.08}_{-0.14} (syst.)}
\newcommand{\fiidiiiful}{F_2^{D(4)}\,(\beta,\,Q^2,\,x,\,t)}
\newcommand{\fiipom}{\tilde F_2^D}
\newcommand{\ALPHA}{1.10\pm0.03 (stat.) \pm0.04 (syst.)}
\newcommand{\ALPHAZ}{1.15\pm0.04 (stat.)^{+0.04}_{-0.07} (syst.)}
\newcommand{\fiipomarg}{\fiipom\,(\beta,\,Q^2)}
\newcommand{\pomflux}{f_{\pom / p}}
\newcommand{\nxpom}{1.19\pm 0.06 (stat.) \pm0.07 (syst.)}
\newcommand {\gapprox}
   {\raisebox{-0.7ex}{$\stackrel {\textstyle>}{\sim}$}}
\newcommand {\lapprox}
   {\raisebox{-0.7ex}{$\stackrel {\textstyle<}{\sim}$}}
\def\gsim{\,\lower.25ex\hbox{$\scriptstyle\sim$}\kern-1.30ex%
\raise 0.55ex\hbox{$\scriptstyle >$}\,}
\def\lsim{\,\lower.25ex\hbox{$\scriptstyle\sim$}\kern-1.30ex%
\raise 0.55ex\hbox{$\scriptstyle <$}\,}
\newcommand{\pomfluxarg}{f_{\pom / p}\,(x_\pom)}
\newcommand{\dsf}{\mbox{$F_2^{D(3)}$}}
\newcommand{\dsfva}{\mbox{$F_2^{D(3)}(\beta,Q^2,x_{I\!\!P})$}}
\newcommand{\dsfvb}{\mbox{$F_2^{D(3)}(\beta,Q^2,x)$}}
\newcommand{\dsfpom}{$F_2^{I\!\!P}$}
\newcommand{\gap}{\stackrel{>}{\sim}}
\newcommand{\lap}{\stackrel{<}{\sim}}
\newcommand{\fem}{$F_2^{em}$}
\newcommand{\tsnmp}{$\tilde{\sigma}_{NC}(e^{\mp})$}
\newcommand{\tsnm}{$\tilde{\sigma}_{NC}(e^-)$}
\newcommand{\tsnp}{$\tilde{\sigma}_{NC}(e^+)$}
\newcommand{\st}{$\star$}
\newcommand{\sst}{$\star \star$}
\newcommand{\ssst}{$\star \star \star$}
\newcommand{\sssst}{$\star \star \star \star$}
\newcommand{\tw}{\theta_W}
\newcommand{\sw}{\sin{\theta_W}}
\newcommand{\cw}{\cos{\theta_W}}
\newcommand{\sww}{\sin^2{\theta_W}}
\newcommand{\cww}{\cos^2{\theta_W}}
\newcommand{\trm}{m_{\perp}}
\newcommand{\trp}{p_{\perp}}
\newcommand{\trmm}{m_{\perp}^2}
\newcommand{\trpp}{p_{\perp}^2}
\newcommand{\alp}{\alpha_s}

\newcommand{\alps}{\alpha_s}
\newcommand{\sqrts}{$\sqrt{s}$}
\newcommand{\LO}{$O(\alpha_s^0)$}
\newcommand{\Oa}{$O(\alpha_s)$}
\newcommand{\Oaa}{$O(\alpha_s^2)$}
\newcommand{\PT}{p_{\perp}}
\newcommand{\JPSI}{J/\psi}
\newcommand{\sh}{\hat{s}}
\newcommand{\uh}{\hat{u}}
\newcommand{\MP}{m_{J/\psi}}
\newcommand{\PO}{I\!\!P}
\newcommand{\xbj}{x}
\newcommand{\xpom}{x_{\PO}}
\newcommand{\ttbs}{\char'134}
\newcommand{\xpomlo}{3\times10^{-4}}  
\newcommand{\xpomup}{0.05}  
\newcommand{\dgr}{^\circ}
\newcommand{\pbarnt}{\,\mbox{{\rm pb$^{-1}$}}}
\newcommand{\gev}{\,\mbox{GeV}}
\newcommand{\gevc}{\,\mbox{GeV/c}}
\newcommand{\mev}{\,\mbox{MeV}}
\newcommand{\WBoson}{\mbox{$W$}}
\newcommand{\fbarn}{\,\mbox{{\rm fb}}}
\newcommand{\fbarnt}{\,\mbox{{\rm fb$^{-1}$}}}
\newcommand{\dsdx}[1]{$d\sigma\!/\!d #1\,$}
\newcommand{\eV}{\mbox{e\hspace{-0.08em}V}}
\newcommand{\nbarn}{\,\mbox{{\rm nb}}}
\newcommand{\nbarnt}{\,\mbox{{\rm nb$^{-1}$}}}
\newcommand{\cm}{\,\mbox{cm}}

%
%
\newcommand{\qsq}{\ensuremath{Q^2} }
\newcommand{\gevsq}{\, \ensuremath{\mbox{GeV}^2} }
\newcommand{\et}{\ensuremath{E_t^*} }
\newcommand{\rap}{\ensuremath{\eta^*} }
\newcommand{\gp}{\ensuremath{\gamma^*}p }
\newcommand{\dsiget}{\ensuremath{{\rm d}\sigma_{ep}/{\rm d}E_t^*} }
\newcommand{\dsigrap}{\ensuremath{{\rm d}\sigma_{ep}/{\rm d}\eta^*} }

\newcommand{\dstar}{\ensuremath{D^*}}
\newcommand{\dstarp}{\ensuremath{D^{*+}}}
\newcommand{\dstarm}{\ensuremath{D^{*-}}}
\newcommand{\dstarpm}{\ensuremath{D^{*\pm}}}
\newcommand{\zDs}{\ensuremath{z(\dstar )}}
\newcommand{\Wgp}{\ensuremath{W_{\gamma p}}}
\newcommand{\ptds}{\ensuremath{p_t(\dstar )}}
\newcommand{\etads}{\ensuremath{\eta(\dstar )}}
\newcommand{\ptj}{\ensuremath{p_t(\mbox{jet})}}
\newcommand{\ptjn}[1]{\ensuremath{p_t(\mbox{jet$_{#1}$})}}
\newcommand{\etaj}{\ensuremath{\eta(\mbox{jet})}}
\newcommand{\detadsj}{\ensuremath{\eta(\dstar )\, \mbox{-}\, \etaj}}

\newcommand{\degree}{^{\circ}}

\def\Journal#1#2#3#4{{#1} {\bf #2} (#3) #4}
\def\NCA{\em Nuovo Cimento}
\def\NIM{\em Nucl. Instrum. Methods}
\def\NIMA{{\em Nucl. Instrum. Methods} {\bf A}}
\def\NPB{{\em Nucl. Phys.}   {\bf B}}
\def\PLB{{\em Phys. Lett.}   {\bf B}}
\def\PRL{\em Phys. Rev. Lett.}
\def\PRD{{\em Phys. Rev.}    {\bf D}}
\def\ZPC{{\em Z. Phys.}      {\bf C}}
\def\EJC{{\em Eur. Phys. J.} {\bf C}}
\def\CPC{\em Comp. Phys. Commun.}

\begin{titlepage}

\noindent
\begin{flushleft}
{\tt DESY 12-072    \hfill    ISSN 0418-9833} \\
{\tt May 2012}                  \\
\end{flushleft}

\noindent

\vspace{2cm}
\begin{center}
\begin{Large}

{\bf Measurement of Beauty Photoproduction near Threshold using Di-electron Events with the H1 Detector at HERA }

\vspace{2cm}

H1 Collaboration

\end{Large}
\end{center}

\vspace{2cm}

\begin{abstract}
The cross section for $ep \rightarrow e\, b\bar{b} X$  in  photoproduction  
is measured with the H1 detector at the $ep$-collider HERA. 
The decay channel $b\bar{b} \rightarrow ee X'$ is selected by identifying the 
semi-electronic decays of the $b$-quarks.
The total production cross section is measured in the kinematic range given by
the photon virtuality $Q^2 \leq 1 \gev^2$, the inelasticity $0.05 \leq y \leq 0.65$ and the 
pseudorapidity of the $b$-quarks $|\eta(b)|, |\eta(\bar{b})|\leq 2$.
The differential production cross section is measured as a function
of the  average transverse momentum of the beauty quarks $\langle P_T(b) \rangle$
down to the threshold. 
The results are compared to next-to-leading-order QCD predictions.
\end{abstract}

\vspace{1.5cm}

\begin{center}
Submitted to \EJC \;\;
\end{center}

\end{titlepage}

%
%
%
\begin{flushleft}

F.D.~Aaron$^{5,45}$,           
C.~Alexa$^{5}$,                
V.~Andreev$^{25}$,             
S.~Backovic$^{30}$,            
A.~Baghdasaryan$^{38}$,        
S.~Baghdasaryan$^{38}$,        
E.~Barrelet$^{29}$,            
W.~Bartel$^{11}$,              
K.~Begzsuren$^{35}$,           
A.~Belousov$^{25}$
% ...


\bigskip{\it
% ...
$ ^{5}$ National Institute for Physics and Nuclear Engineering (NIPNE) ,
          Bucharest, Romania$^{ k}$ \\
% ...
$ ^{11}$ DESY, Hamburg, Germany \\
% ...
$ ^{25}$ Lebedev Physical Institute, Moscow, Russia \\
% ...
$ ^{29}$ LPNHE, Universit\'e Pierre et Marie Curie Paris 6,
          Universit\'e Denis Diderot Paris 7, CNRS/IN2P3, Paris, France \\
 $ ^{30}$ Faculty of Science, University of Montenegro,
          Podgorica, Montenegro$^{ l}$ \\
% ...
$ ^{35}$ Institute of Physics and Technology of the Mongolian
          Academy of Sciences, Ulaanbaatar, Mongolia \\
% ...
$ ^{38}$ Yerevan Physics Institute, Yerevan, Armenia \\
% ...
$ ^{45}$ Also at Faculty of Physics, University of Bucharest,
          Bucharest, Romania \\
% ...
$ ^b$ Supported by the UK Science and Technology Facilities Council,
      and formerly by the UK Particle Physics and
      Astronomy Research Council \\
% ...
$ ^k$ Supported by the Romanian National Authority for Scientific Research
      under the contract PN 09370101 \\
 $ ^l$ Partially Supported by Ministry of Science of Montenegro,
      no. 05-1/3-3352 \\
}
\end{flushleft}
%

\newpage

\section{Introduction}
\begin{figure}[hhh]
\center
\setlength{\unitlength}{4cm}
\epsfig{file=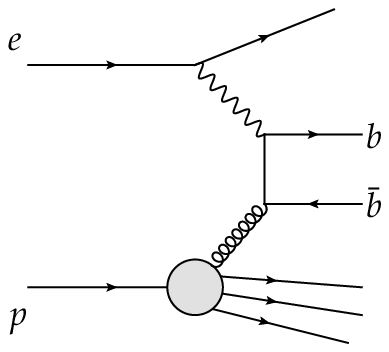,width=4cm} 
\put(-1.05,1.1) {a)}
\hspace{2cm}
\epsfig{file=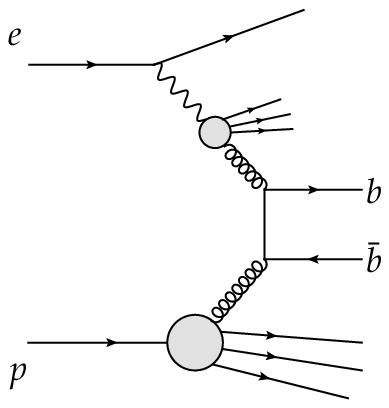,width=4cm} 
\put(-1.0,1.1) {b)}
\caption{Generic leading order diagrams for $b\bar{b}$ production in $ep$ collisions.
The diagram a) is referred to as direct or pointlike, the diagram b) is referred to as 
resolved or hadronlike.}
\label{fig: FeynmanDiag} 
\end{figure}

In $ep$ collisions at HERA beauty quarks are mainly produced as
$b\bar{b}$ pairs via the fusion of a quasi-real photon emitted by the incoming
electron (or positron)
and a gluon of the proton as depicted in figure~\ref{fig: FeynmanDiag}a. 
This process is referred to as direct or pointlike and 
can be calculated using perturbative quantum chromodynamics (QCD) due to the 
large scale provided by the mass of the heavy $b$-quark and the correspondingly 
small coupling  $\alpha_s$. Resolved processes where the photon fluctuates into a 
hadronic state before undergoing a hard collision, as indicated in 
figure~\ref{fig: FeynmanDiag}b, are expected to be largely suppressed compared 
to the direct production process, because of the large $b$-quark mass.
Due to the dominance of the direct process over
the resolved process, the production of $b$-quarks in $ep$ collisions
at HERA is an excellent testing ground for QCD predictions.

Theory uncertainties in the prediction of the cross section, which are mainly
related to the renormalisation and factorisation scales, are expected to be smaller
for beauty production than for charm production.
The study of beauty photoproduction near threshold is of particular theoretical interest 
as the only hard scale in this process is provided by the $b$-quark mass, and other scales 
like the photon virtuality ($Q^2 \approx 0 \gev^2$ in photoproduction) or
the transverse momentum of the $b$-quark can be neglected.

At HERA the beauty cross section in photoproduction $ep \rightarrow e\, b\bar{b} X$ 
has been measured by the
H1~\cite{Mira, Aktas:2006vs,  Aktas:2005bt, Aktas:2005zc, Adloff:1999nr}
 and 
ZEUS~\cite{ZEUS:2011aa, Chekanov:2008tx, Chekanov:2008zz, Chekanov:2008aaa, Chekanov:2006sg,  
Chekanov:2003si, Breitweg:2000nz}
collaborations and compared to calculations~\cite{Fixone97, massive, Frixione:1994dv}
at next-to-leading order (NLO) QCD, performed in the  fixed flavour
number scheme in which the beauty quark is treated as massive.
In general the predictions using the  factorisation and renormalisation scale 
$\mu_R = \mu_F = \sqrt{m_b^2 + P_T(b)^2}$ do not agree well with the data.
In particular at low values of the transverse momentum of the beauty quarks
$P_T(b) \approx 0\gev$, 
i.e. in the phase space region where the only hard scale involved is the
$b$-quark mass $m_b$, the measurements show a tendency to lie above the prediction.
The choice of a lower scale, $\mu_R = \mu_F = 1/2 \,\sqrt{m_b^2 + P_T(b)^2}$, 
leads to a better agreement of the prediction with the data~\cite{Geiser:2008zz}.

In the present analysis a measurement of the differential
beauty cross section at HERA in photoproduction as function 
of the quadratically averaged transverse momentum of the produced beauty quarks,
${\mathrm{d}\sigma}/{\mathrm{d}\langle P_T(b) \rangle}$,
is made down to the $b \bar{b}$-production threshold, using a novel technique 
based on low momentum electron identification.

Most of the previous beauty measurements at HERA in photoproduction and  
deep-inelastic scattering (DIS, $Q^2 \gtrsim 1\gevsq$) identified jets of $b$-quarks 
using single leptons tags~\cite{Mira, Abramowicz:2011kj, Chekanov:2008tx, Chekanov:2009kj, Chekanov:2008aaa, 
Chekanov:2006sg, Chekanov:2003si, Adloff:1999nr,Aktas:2005zc, Abramowicz:2010zq, Chekanov:2004tk} 
or displaced vertices~\cite{Aktas:2006vs, :2009ut, Aaron:2010ib, Aktas:2005iw,
Aktas:2004az, ZEUS:2011aa}.
Jet-based $b$-tagging algorithms are in general very efficient at high transverse momenta 
of $b$-quarks, $P_T(b)>6-7\gev$, but degrade significantly for lower values due to the 
absence of the boost and the short decay length. 
In addition  $b$-quarks  almost at rest lead to isotropic decay topologies of the final 
state where jet finders usually fail. A second class of analyses used double tags to 
select $b \bar{b}$ pairs either by reconstructing  two muons~\cite{Chekanov:2008zz} 
or a muon and a  $D^*$ meson~\cite{Aktas:2005bt, Chekanov:2006sg}, utilizing the 
semileptonic decay channel $b\rightarrow \mu X'$ and the decay channel $b \rightarrow D^* X'$, 
respectively. Lower values of the transverse momentum of the $b$-quarks become 
accessible by the use of lepton tags without requiring jets, where the minimum 
$P_T(b)$ value is determined by the minimum transverse momentum cut on the lepton. 
For muons this cut is typically at $P_T(\mu) \approx 2\gev$, and therefore too high to 
measure efficiently the production cross sections of $b$-quarks near threshold.
 
In the present analysis the differential beauty cross section is measured using
electron pairs, exploiting the double-semileptonic decay $b \bar{b} \rightarrow e e X'$, 
with online and offline $P_T(e)$~thresholds for the electron identification of about $1\gev$.
The events were recorded by identifying low
momentum electrons already online using a dedicated trigger, which
recorded data in the year 2007 with a corresponding integrated luminosity of $48.1\pbarnt$. 
This low cut on the transverse electron momentum, $P_T(e)$, improves not only the
total acceptance but also makes the low $P_T(b)$ phase space experimentally accessible.

\section{Monte Carlo Simulations and QCD Calculations}
\label{sec: MC_FMNR}

The Monte Carlo generators PYTHIA~\cite{Sjostrand:2001yu} and
 CASCADE~\cite{Jung:2000hk} are used to determine the signal efficiency and
 the detector acceptance for the
 process $ep \rightarrow e\, b \bar{b} X \rightarrow e\, ee X'$, and to simulate 
 the production of charm quarks.
Differences in the predictions are taken into account as systematic
uncertainty, see section~\ref{sec: Cross Section Determination and 
Systematic Uncertainties}.  For the production of $\JPSI$ mesons only CASCADE 
and for the production of light quarks in photoproduction  only PYTHIA is used.
Deep-inelastic scattering is simulated using the Monte Carlo generator 
RAPGAP~\cite{RAPGAP}.

In PYTHIA leading order matrix elements are implemented taking into account the 
mass of the heavy quarks. The CTEQ6L~\cite{CTEQ6} set of proton parton density 
functions is used. The parton shower evolution in PYTHIA is based on the DGLAP 
equations~\cite{Gribov:ri}. In addition to the direct process, the resolved photon 
component is calculated by using the photon parton density 
function \mbox{SAS 2D}~\cite{Sjostrand:1995yc}.

For the CASCADE simulation the direct $\gamma^* p \rightarrow b\bar{b}$ and 
$\gamma^* p \rightarrow  c\bar{c}$ processes are implemented using off-shell 
matrix elements, which are convoluted with $k_T$-unintegrated\footnote{$k_T$ 
denotes the transverse momentum of the parton.}  proton parton density 
functions. The A0~\cite{Jung:2004gs} set of parton density functions is used.
The parton evolution in CASCADE is based on the  CCFM evolution equation~\cite{ccfm} 
for the initial state parton shower.

In PYTHIA, CASCADE and RAPGAP higher order QCD corrections
are included by simulating parton showers in the initial and final state.
These  Monte Carlo generators use the Lund String Model~\cite{Lund} 
for simulating the hadronisation of light quarks. For the hadronisation of heavy quarks 
the Bowler fragmentation model~\cite{Bowler}  is employed with parameters as used 
in a previous analysis~\cite{Aaron:2010ib}.

In order to correct for detector effects and to estimate the systematic uncertainties 
associated with the measurement, the generated events are passed through a 
detailed simulation of the detector response based on the GEANT 
program~\cite{Brun:1987ma} and through the same reconstruction and analysis 
software as is used for the data.

Theory cross sections  are calculated in NLO QCD in the fixed flavour number scheme 
using the program FMNR~\cite{Fixone97, massive, Frixione:1994dv} in order to compare 
with the data. These calculations are expected to give reliable results in the kinematic 
region considered here, where the transverse momentum of the heavy quark is of the 
same order of magnitude as its mass. The calculations are performed  as a function of 
the quadratically averaged transverse momentum of the produced beauty pair  
\begin{linenomath*}
\begin{equation}
\label{eq: DefPtB}
\langle P_T(b) \rangle = \sqrt{(P_{T,b}^2 + P_{T,\bar{b}}^2)/2}   \quad .
\end{equation}
\end{linenomath*}
The prediction of FMNR is evaluated for the direct and resolved photon processes.  
For the proton the CTEQ6M~\cite{CTEQ6} set and for the photon the GRV-HO~\cite{Gluck1998} 
set of parton density functions are used. In this analysis,
the renormalisation and factorisation scales are chosen to be equal,
$\mu_R = \mu_F = \mu_0$, with $\mu_0 = 1/2 \, \sqrt{m_b^2 + \langle P_T(b) \rangle  ^ 2}$ 
and $m_b = 4.75 \gev$. 
The  value used for the QCD scale $\Lambda_{\mathrm{QCD}}$ corresponds to 
the value of the strong coupling constant $\alpha_s(M_{\mathrm{Z}})= 0.118$.
The theoretical uncertainty of the prediction is evaluated by varying 
the scales $\mu_R$ and $\mu_F$ simultaneously
in the window $\mu_0 / 2 < \mu_{R,F} < 2 \mu_0$ and the  beauty mass in 
the range $4.5 < m_b < 5.0 \gev$. 
By recalculating the cross section with different parton density functions
the theoretical uncertainty due to the choice of the photon and proton parton 
density functions is found to be much smaller than
the theoretical uncertainties and thus is neglected.

\section{H1 Detector}
A detailed description of the H1 detector can be found elsewhere \cite{H1det, Appuhn:1996na}. 
In the following, only  detector components relevant to this analysis are 
briefly discussed. The origin of the H1 coordinate system is the nominal 
$ep$~interaction point, with the direction of the proton
beam defining the positive $z$~axis (forward direction). 
Transverse momenta are measured in the \mbox{$x$-$y$}~plane. Polar 
($\vartheta$) and azimuthal ($\phi$) angles are measured with respect to this 
reference system. The  pseudorapidity is defined to be  $\eta = - \ln \tan(\vartheta/2)$.

In the central region (\mbox{$15^\circ\!<\!\vartheta\!<\!165^\circ$}) the
interaction point is surrounded by the central tracking detector (CTD).
The CTD comprises two large cylindrical jet chambers
(CJC1 and CJC2) and the silicon vertex detector~\cite{pitzl}.
The CJCs are separated by a drift chamber
which improves the $z$~coordinate reconstruction.
The CTD detectors are arranged concentrically around the interaction 
region in a solenoidal magnetic field of \mbox{$1.16\ {\rm T}$}. The 
trajectories of the charged particles are measured with a transverse 
momentum resolution of $\sigma(p_T)/p_T \approx 0.2\% \, p_T/{\rm GeV} 
\oplus 1.5\%$.
In addition the CJCs provide a measurement of
the specific ionisation energy loss $\mathrm{d}E/\mathrm{d}x$ of charged particles
with a relative resolution of $6.5\%$ for long tracks.
A set of five cylindrical multiwire proportional chambers~\cite{cip2000} mainly
used for first level triggering are situated inside the inner CJC1 covering 
the polar angular region $11^\circ < \vartheta < 169^\circ$.
The forward tracking detector and the backward proportional chamber 
measure tracks of charged particles at smaller
\mbox{($7^\circ\!<\!\vartheta\!<\!25^\circ$)} 
and larger \mbox{($155^\circ\!<\!\vartheta\!<\!175^\circ$)} polar angles
than the central tracker, respectively.

The liquid argon (LAr) sampling calorimeter~\cite{Andrieu:1993kh} surrounds the tracking
chambers and has a  polar angle coverage of 
\mbox{$4^\circ\!<\!\vartheta\!<\!154^\circ$}.
It consists of an inner electromagnetic section with lead absorbers and an outer 
hadronic section with steel absorbers.
The LAr calorimeter is divided into eight wheels  along the beam axis. The electromagnetic and 
the hadronic sections are highly segmented in the transverse and the longitudinal directions.
Energies of electromagnetic showers are measured with a precision of
$\sigma(E)/E=12\%/\sqrt{E/\gev}\oplus 1\%$ and energies of hadronic showers with
$\sigma(E)/E=50\%/\sqrt{E/\gev}\oplus 2\%$, as determined in test beam experiments~\cite{Andrieu:1994yn,h1testbeam}. 
In the backward
region (\mbox{$153^\circ\!<\!\vartheta\!<\!178^\circ$}), particle energies are measured
by a lead-scintillating fibre calorimeter (SpaCal)~\cite{Appuhn:1996na}. 
\par 
The luminosity is determined from the rate of the elastic QED Compton
process $ep \rightarrow e\, \gamma p$, with the electron and the photon detected
in the SpaCal~calorimeter, and the rate of DIS events measured in the SpaCal~calorimeter~\cite{QEDComptonAnalysis}.

For data collection a four level trigger system is employed, of which the
first two levels are implemented in hardware. The first level trigger (L1)
is based on various sub-detector components, which are combined and refined at
the second level (L2). 
The third level (L3) is a software based trigger using combined
L1 and L2  trigger information from various subdetector components. 
Fully reconstructed events are subject to an
additional selection at the software filter farm (L4).

The data used for this measurement were recorded by the Fast Track Trigger 
(FTT)~\cite{FTT, Jung2007a} which, based on hit information provided by the CJCs,
reconstructs tracks with subsequently refined granularity at the first two trigger levels,
first in the \mbox{$x$-$y$}~plane at L1 and then in three dimensions at L2. 
Of special importance is the third trigger level integrated in the FTT~\cite{Jung2007a},  
which identifies low energy electrons ($E > 1\gev$)~\cite{lcaminada,msauter} by 
combining FTT~tracks with energy depositions reconstructed in the LAr calorimeter by the 
Jet Trigger (JT)~\cite{Olivier:2011zz}.

\section{Experimental Method}
The data sample used for this analysis was recorded in the year 2007, when
positrons at an energy of $27.6\gev$  collided with protons at
$920\gev$, and
when all trigger levels of the FTT and the JT were in operation. 
The recorded data corresponds to a total integrated luminosity of $48.1\pbarnt$.

In this analysis the measurement of $b\bar{b}$ photoproduction is  based on
the identification and selection of two electrons
in the LAr calorimeter at low transverse momentum  
$P_{T}(e) >  1\gev$ to tag the semi-electronic decays of the $b$-quarks. 
In about $2\%$ of all $b\bar{b}$-decays  two electrons originate from the
same $b$-quark from the decay chain $b\rightarrow c\; e^-
\bar{\nu_e} \rightarrow s\; e^- e^+ \bar{\nu_e} \nu_e$. 
In about $4\%$ of all $b\bar{b}$-decays  the two electrons originate 
from decays of different $b$-quarks, where they are either produced directly in the
semi-electronic $b$-decays or in the subsequent semi-electronic $c$-decays. 
The electrons can be either of opposite charge (combinations
$b\bar{b}\rightarrow e^+e^- X$ and $c\bar{c}\rightarrow e^+e^- X$)
or of same charge (combinations
$b\bar{c}\rightarrow e^-e^- X$ and $\bar{b}c\rightarrow e^+e^+ X$).
These charge relations hold only in the case of no $B \bar{B}$~mixing. 
In the following all possible combinations 
including $B \bar{B}$~mixing
are considered in order to discriminate
$b\bar{b}$ decays against semi-electronic decays of  $c\bar{c}$ events. 
Electron pairs from $\JPSI$ decays are distinguished from those 
from $b$-decays by reconstructing their invariant mass.
Misidentified electrons originating mainly from the light quark background are
constrained by varying the cuts on the electron identification described in 
section~\ref{sec: DataAnalysis}.

\subsection{Online Electron Identification}
Events containing several  tracks and one or two electron candidates
compatible with the signature of semi-electronic $b$-decays are triggered,
using the FTT on the trigger levels L1 to L3.
On the first trigger level more than five tracks with transverse
momentum thresholds in the range $0.1 - 1.8~\gev$ are required. 
These high multiplicities  are verified at the second trigger
level, exploiting the higher track resolution available at this level. 
On the third trigger level the track information as determined by FTT-L2 
is combined with the energy depositions as measured in the LAr calorimeter by the
Jet Trigger~\cite{lcaminada, msauter, Jung2007a} to identify electrons.
Electron candidates are required to fulfill a geometrical track-cluster matching condition
using the distance variables 
$\Delta \vartheta = | \vartheta_{\mathrm{FTT}} - \vartheta_{\mathrm{JT}}|$ and
$\Delta \varphi = | \varphi_{\mathrm{FTT}} - \varphi_{\mathrm{JT}}|$.
In addition the  transverse  momentum $P_{T, \mathrm{FTT}}$ as measured with the FTT-L2  
has to be compatible with the
associated transverse energy $E_{T, \mathrm{JT}}$ measured in the LAr calorimeter by the JT. 
A lower cut on the quantity $E_{T, \mathrm{JT}}/P_{T, \mathrm{FTT}}$ is 
used to discriminate electrons against hadrons, which deposit significantly 
less energy in the non-compensating LAr calorimeter.

\renewcommand{\arraystretch}{1.0} 
\begin{table}[tt]
\begin{center}
\begin{tabular}{|l|c|c|c|c|c|c|}
\hline
	Subtrigger & 					 
	\# ele. cand. &  $P_{T,\mathrm{FTT}}$ [\mbox{GeV}]  &
	\raisebox{-0.5ex}{$ \frac{E_{T,\mathrm{JT}}}{P_{T,\mathrm{FTT}}}$ [\%]} &
	$\Delta \varphi$ [rad] &
	$\Delta \vartheta$ [rad]& $\mathcal{L}$ [$\mathrm{pb^{-1}}$]\\ 
\hline \hline
low-$P_T$		& $\geq2$	&$>1.2$ 	&	$>30$	& $<0.30$	& $<0.25$	& $25.1$\\
medium-$P_T$	& $\geq1$	&$>1.5$ 	&	$>50$	& $<0.15$	& $<0.20$	& $13.1$\\
high-$P_T$		& $\geq1$	&$>2.0$	&	$>60$ & $<0.20$		& $<0.20$	& $33.5$\\
\hline

\end{tabular}
\caption{L3-online cuts used to trigger electron candidates.
Explanations to the cuts are given in the text.
The last column contains the prescale corrected integrated luminosity of each subtrigger.
The medium-$P_T$ subtrigger  was commissioned at a later stage.}
\label{tab:EleTrig}
\end{center}
\end{table}

For this analysis three subtriggers are used, which have identical L1 and similar L2 trigger 
conditions, but different conditions on L3 as summarized in table~\ref{tab:EleTrig}. 
The subtrigger with the lowest transverse momentum threshold of 
$P_{T,\mathrm{FTT}}>1.2~\gev$ requires events with at least two electron 
candidates. The other two subtriggers select events with a minimum of one electron candidate 
with $P_{T, \mathrm{FTT}}$ thresholds of  $1.5$ and $2.0~\gev$.
The three data sets recorded by these FTT-JT based subtriggers cover an overlapping kinematic 
phase space, but correspond to different integrated luminosities due to different trigger 
prescale factors. The three data sets are combined using a weighting method~\cite{msauter} 
to account for correlated triggers with prescales. The individual prescale corrected
luminosities are also given in table~\ref{tab:EleTrig}.

\subsection{Offline Electron Identification}
\label{sec: OfflineEleID}
Electrons in the polar angle range of $20 \degree <\vartheta(e)<140 \degree$
and with a transverse momentum of $P_T(e)>1\gev$, with $P_T(e)$ and $\vartheta(e)$ 
measured from the electron track, are identified  using  energy depositions in the 
LAr calorimeter and  specific ionisation loss measured in the CJCs.
Two estimators, $D_{\mathrm{calo}}$ and $D_{\mathrm{d}E/\mathrm{d}x}$,  are defined 
to discriminate electrons from background. The background, which is mainly due to  pions 
misidentified as electrons and to a lesser extent due to kaons and anti-protons, is largely 
suppressed by combining the two independent estimators into  a combined estimator 
$D_{\mathrm{ele}}$, as explained in appendix~\ref{Appendix Electron Discriminator Combination}.
The three estimators are defined such that $D=1$ for genuine electrons and 
$D=0$ for pion background.

The calorimeter based electron identification~\cite{msauter} is track seeded, 
which means the cluster shape estimators are calculated from energy
deposits in LAr calorimeter cells lying
within a cylinder of $30\cm$
around the extrapolated track trajectory. The Cluster energies are corrected for energy losses in 
the dead material in front of the LAr calorimeter.
Electron candidates with energy depositions close to inactive regions between LAr calorimeter 
modules are rejected. Five estimators are defined: 
four cluster shape variables and the ratio of the energy deposited in the electromagnetic part of the
LAr calorimeter to the momentum  of the corresponding track.
These  estimators together with the logarithm of the total energy and the $z$~position of the cluster,
are mapped onto one single estimator $D_{\mathrm{calo}}$ using the artificial neural network
Multilayer Perceptron~\cite{TMVA}.

The measured specific ionisation loss of the track, $\mathrm{d}E/\mathrm{d}x$, is translated into 
$\chi^2$-probabilities of corresponding particle hypotheses
$P(\mathrm{d}E/\mathrm{d}x, e)$ for electrons and $P(\mathrm{d}E/\mathrm{d}x, \pi)$ for pions, 
which constitute the main background. 
From both probabilities the estimator  
\begin{linenomath*}
\begin{equation}
D_{\mathrm{d}E/\mathrm{d}x} = \frac{P(\mathrm{d}E/\mathrm{d}x, e)}{ P(\mathrm{d}E/\mathrm{d}x, e) + P(\mathrm{d}E/\mathrm{d}x, \pi)}
\end{equation}
\end{linenomath*}
is constructed.
The simulation of the specific ionisation was studied in detail in order to 
describe precisely the measured energy losses \cite{ehennekemper}.

The performance of both discriminator variables is validated using Monte Carlo
and data
samples of identified electrons and pions
in the transverse momentum range of interest, $1< P_T(e) < 5\gev$,
selected in decays $\JPSI \rightarrow e^+e^-$ and $K_s^0 \rightarrow \pi^+ \pi^-$, 
by means of the ``tag and probe method''~\cite{msauter}. 

The simulation describes  well the distribution of the
discriminators $D_{\mathrm{calo}}$ and $D_{\mathrm{d}E/\mathrm{d}x}$ as 
measured in data, as can be seen in figure~\ref{fig:ElePiDiscriminator}.
The deviations of the simulation from the data at small $D$ values in the electron sample
are due to a small remaining pion contamination in the data.
Also the combined estimator $D_{\mathrm{ele}}$ is found to be well described by the
simulation and shows an excellent separation  
of the electron signal from the pion background. 
Isolated electrons are selected for $D_{\mathrm{ele}} = 0.825$ with an efficiency 
of more than $90\%$ for a pion background rejection of about $99\%$.

\subsection{Event Selection}
A di-electron sample is obtained by selecting events with
two or more offline reconstructed electron candidates, requiring 
$D_{\mathrm{ele}}>0.825$.
To account for the $P_T$~resolution of the third trigger level, the
$P_T$~cut on electron tracks reconstructed offline is raised by 
$100\mev$  above the  trigger 
threshold of the respective subtrigger (see table~\ref{tab:EleTrig}), which recorded the 
event.
This implies two electrons with $P_T>1.3\gev$ for the low-$P_T$ subtrigger and
one electron with $P_T>1.6\gev$ ($P_T>2.1\gev$) for the medium-$P_T$ 
(high-$P_T$) subtrigger.

In order to remove background from non-$ep$ sources, the measurement
of a good event vertex is required. The event vertex  
is reconstructed from all charged tracks of an event and its position along the beamline 
has to be within $30\cm$ around the nominal interaction point.
In addition, timing vetoes are applied to further reduce non-$ep$ interaction 
induced backgrounds.

In order to reject background from DIS, events with
a positron in the LAr calorimeter identified 
by the standard electron 
identification~\cite{ElectronFindersLAR} and with $E(e^+)>8\gev$ are rejected.
As the $P_T(e)$-distribution of semileptonically decaying $b$-quarks falls steeply,
almost all $b$-decay positrons are at low energies 
and  thus not affected by this cut.
Also DIS events are rejected which have an electromagnetic cluster in the 
SpaCal~calorimeter with energy above $8\gev$
consistent with originating from the scattered beam positron. 
Events with $Q^2 \lesssim 2 \gevsq$ are not rejected by these cuts, since the 
beam positrons leave the detector undetected along the beam pipe.

Only events with measured inelasticities in the phase space region of this measurement, 
$0.05 < y_{\mathrm{h}} < 0.65$, are
accepted. 
The inelasticity variable is reconstructed from the sum over all  final state particles 
$y_{\mathrm{h}} = \sum_{i} (E_i - P_{z, i}) / (2  E_{\mathrm{e^+ beam}})$, where $E_{\mathrm{e^+ beam}}$ denotes the
energy of the beam positron.
Particles belonging to the
hadronic final state (HFS) are reconstructed using a combination of tracks and calorimeter deposits in an energy flow
algorithm that avoids double counting~\cite{peezportheault}.
 $E_i$ and $P_{z, i}$ denote the
energies and longitudinal momenta of all final state particles, 
which correspond to the visible hadronic final state in case of photoproduction, and
in case of DIS background also includes the scattered positron. 
The upper cut on the inelasticity suppresses effectively remaining DIS events.

The beauty signal is further enriched by rejecting electron candidates, which are
in a dense hadronic environment. 
For this purpose the variable $R_{\mathrm{E,cone}}$  is defined as
the ratio of the summed  energy of all HFS particles in a cone of $18 \degree$ around the electron
track direction, $E_{\mathrm{cone}}$, to
the electron energy $E_{e}$, which must not exceed an upper threshold:
\begin{linenomath*}
\begin{equation}
R_{\mathrm{E,cone}} = \frac{E_{\mathrm{cone}}}{E_{e}} < 350 \%  \quad .
\end{equation}
\end{linenomath*}
The effect of this cut is twofold: First, it reduces
misidentified electron candidates resulting from 
overlapping showers in the LAr calorimeter. Second, it 
enriches electrons from semileptonic beauty
decays, which are in general  isolated from hadrons
due to the large $b$-mass. 

Finally, electrons from photon conversions are rejected by the three 
following cuts.
First, the distance of closest approach in the transverse plane, $\mathrm{dca}_e$, of 
the electron tracks to the beam line is restricted to be smaller than $0.2\cm$. 
Second, a photon conversion finder searching for displaced
vertices is used to identify electrons originating from the photon conversion
process $\gamma \rightarrow e^+ e^-$.
Third, the invariant mass of 
the selected electron pairs is required to be  $m_{e1,e2} > 1.2\gev$. This
cut rejects $e^+e^-$ pairs from Dalitz decays and most of the remaining background from
photon conversions.

The selection cuts are summarised in table~\ref{tab: selection}. After applying all cuts 
about $1500$ electron pairs are selected. In the rare cases with more than two selected 
electrons per event all pair combinations are considered in the analysis.

\begin{table}[tt]
\begin{center}
\begin{tabular}{|l|}
\hline
\multicolumn{1}{|c|}{\bf \boldmath Overview of the Selection Cuts  }\\
\hline \hline
 Trigger selection \\
\hline
 \hspace{0.5cm} $\bullet$ track multiplicity cuts  \\
 \hspace{0.5cm} $\bullet$ 1 or 2 online identified electrons \\
\hline
\hline 
	Offline electron selection\\
	 \hline
 \hspace{0.5cm} $\bullet$ 2 electron candidates with: \\
 \hspace{1cm} $-$ $D_{\mathrm{ele}} > 0.825$, $R_{\mathrm{E,cone}} < 350 \%$ \\
 \hspace{1cm} $-$ $P_T(e) > 1\gev$, 	$20 \degree < \vartheta(e) < 140 \degree$\\
 \hspace{1cm} $-$ verification of the L3 $P_T(e)$-thresholds $100\mev$ above the $P_T(e)$-threshold of the \\
 \hspace{1.5cm}  respective subtrigger which recorded the event (see table~\ref{tab:EleTrig} and text)\\
\hline \hline
	 Background rejection and further cuts	\\	
\hline
{Rejection of non ep-background:} \\
  \hspace{0.5cm} $\bullet$ good vertex, timing vetoes\\
{Rejection of DIS events:} \\
  \hspace{0.5cm} $\bullet$  no identified scattered beam positron\\
  \hspace{0.5cm} $\bullet$ $0.05 < y_{\mathrm{h}} < 0.65$\\ 
{Rejection of photon conversions and Dalitz decays:}\\
  \hspace{0.5cm} $\bullet$ $m_{e1,e2} > 1.2 \gev$\\
  \hspace{0.5cm} $\bullet$ no converted photon\\
  \hspace{0.5cm} $\bullet$ $\mathrm{dca}_e < 0.2\cm$ \\
\hline
\end{tabular} 
\caption{Overview of the online and offline selection cuts. More details on the
selection procedure can be found in~\cite{msauter}. \label{tab: selection}}
\end{center}
\end{table}

\section{Data Analysis}
\label{sec: DataAnalysis}
The selected di-electron sample is dominated by events from inelastic
$\JPSI$-meson production. While decays of $\JPSI$-mesons can be easily
identified by kinematic reconstruction of the $\JPSI$ mass peak, the separation of
the $b\bar{b}$ signal events from the other backgrounds originating from the
production of light quarks and open charm production is more difficult.

In the following,  the reconstruction of the transverse momentum
of the produced $b$-quarks  and
the flavour separation of the different processes are described.

\subsection{Reconstruction of \boldmath{$b$}-quarks}
The transverse momentum of $b$-quarks is reconstructed for the measurement of 
the differential cross section $\mathrm{d} \sigma / \mathrm{d} \langle P_T(b) \rangle$,
where $\langle P_T(b) \rangle$ is the quadratically averaged transverse momentum
of the $b$ and $\bar{b}$ quark as defined  in equation~\ref{eq: DefPtB}.
The $b\bar{b}$ cross section is largest at small transverse momentum at 
$\langle P_T(b) \rangle \approx m_b$, a kinematic region where standard jet finders cannot 
be used due to isotropic decay topologies.
Therefore an alternative, referenced as the hemisphere method, is exploited. This method was
applied in a previous analysis~\cite{CharmFragmPaper} to reconstruct the directions and
momenta of charm quarks in the production of $c\bar{c}$-pairs in DIS, and 
is also well suited to reconstruct
the transverse momenta of $b$-quarks in $b\bar{b}$ production~\cite{msauter}.  

As illustrated in figure~\ref{Fig:ThrustAxisSchematic}, 
an event is divided into  hemispheres, using
the thrust-axis which is calculated in the laboratory  frame in the plane transverse
to the beam directions (\mbox{$x$-$y$}~plane). 
Using the transverse momenta from all particles of the HFS, the thrust-axis in
the transverse plane is given  by the vector $\vec{a}$ maximising the sum of the projected
transverse momenta onto it, 
\begin{linenomath*}
\begin{equation}
\label{eq: Thrust}
 T = \mathrm{max} (\vec{a})  \Bigg(\dfrac{\sum_{i \in \mathrm{HFS}} |\vec{a} \cdot
   \vec{P}_{T, i} |} {\sum_{i \in \mathrm{HFS}} |\vec{P}_{T, i} | } \Bigg) \quad \quad
\textrm{with} \ \vert\vec{a}\vert=1 \quad .
\end{equation}
\end{linenomath*}
A plane perpendicular to the thrust-axis defines two hemispheres, one of them
containing the fragmentation products of the $b$-quark, and the other one
containing the fragmentation products of the $\bar{b}$-quark. 

Two observables $\vec{P}_{T, \mathrm{hem. I}}$ and $\vec{P}_{T, \mathrm{hem. II}}$ are used 
to reconstruct the mean transverse momentum of the $b$ ($\bar{b}$) quark produced in
the hard interaction. 
These observables, which are derived from the  HFS particles assigned to the
corresponding hemispheres, show a good correlation to the transverse momentum
of the $b$ ($\bar{b}$) quarks in the hard process.
However, the hadronic final state also contains  particles from the so called
proton remnant, leaving the interaction in the positive $z$-direction of the
detector and thus deteriorating the above correlation.
Simulation studies show that the correlation with the $b$-quark transverse
momentum is improved by excluding particles in the
forward direction at polar angles below 15 degrees. The transverse momenta of
the $b$ ($\bar{b}$)-quarks are therefore approximated by:
\begin{linenomath*}
\begin{equation}
\vec{P}_{T,\mathrm{hem.I \ (hem.II)}} = \sum_{\substack{i \ \in \ \mathrm{hem.I}
 \\ ( i \ \in \ \mathrm{ hem.II)}}} \vec{P}_{T, i} 
\quad \quad \textrm{with} \ \vartheta_i>15^\circ \quad .
\end{equation}
\end{linenomath*}

This reconstruction method is very reliable at large $P_{T,b}$, where two hard jets 
are measured in the final state. 
At small $P_{T}(b)$ the transverse momenta of HFS particles in the hemispheres are 
mainly generated by the $b$ and $\bar{b}$-hadron decays  themselves
and are related to the mass of the $b$-quark: 
$|\vec{P}_{T, \mathrm{hem.I}}| \approx |\vec{P}_{T, \mathrm{hem.II}}|
\approx m_b$.
In order to allow for a good reconstruction of $P_{T}(b)$ down to  the $b\bar{b}$
production threshold, i.e. $P_{T}(b) \approx 0\gev$,
the average transverse beauty mass  is used:
\begin{linenomath*}
\begin{align}
\label{eq: M_T}
m_T(b)  = \sqrt{ m_b^2 + \langle P_T(b) \rangle^2}  \quad.
\end{align}
\end{linenomath*}
Detailed studies~\cite{msauter} demonstrated that the average transverse beauty mass can be
well reconstructed from the experimental observables $\vec{P}_{T, \mathrm{hem.I}}$ and
$\vec{P}_{T, \mathrm{hem.II}}$ using the relation:
\begin{linenomath*}
\begin{align}
\label{eq: M_T_est}
m_{T, \mathrm{rec}}(b) &=  \alpha * (|\vec{P}_{T,\mathrm{hem.I}}| + 
|\vec{P}_{T, \mathrm{hem.II}}|)/2 \quad , 
\end{align}
\end{linenomath*}
with $\alpha$ being a constant parameter set to $\alpha = 1.09$, such that the
correlation between generated and reconstructed $m_T(b)$ is maximised.
This correlation
as obtained by simulation is shown in figure~\ref{Fig:ThrustAxis}. For values of 
$m_b$  it in the range $4.5 < m_b < 5.0 \gev$ 
the dependence of this correlation on $m_b$ is negligible.

\subsection{Quark Flavour Separation}

For the discrimination of the $b\bar{b}$ signal against remaining background from
misidentified electrons and for the separation of the different 
quark-flavour components contributing to the di-electron signature, a
template method is used. 
Several independent phase space regions are defined such  that individual
background sources are enhanced in certain regions of the phase space and
can be tested while other contributions are suppressed. 
Finally the $b$-signal (``beauty'') and the background contributions
are obtained by an unfolding procedure. 
Background sources determined by this method are the production of light
quarks (``uds''), 
open charm production (``charm'') and the production of $\JPSI$-mesons (``$\JPSI$'').
The uds background contains also a small fraction of charm and
beauty events, where at least one electron candidate does not originate
from a semi-electronic heavy quark decay.

\subsubsection{Fraction of Light Quarks}
\label{sec: Fraction of Light Quarks}
In order to determine the  background contributions due to
misidentified electrons 
the data sample is grouped in four regions $B1$, $B2$, $B3$ and $S$ using 
different electron quality criteria on  
$D_{\mathrm{ele}}^{\mathrm{min}(e1, e2)}$ and 
$R_{\mathrm{E,cone}}^{\mathrm{max}(e1, e2)}$,
see table~\ref{tab:udsTemplatesDef}.
$D_{\mathrm{ele}}^{\mathrm{min}(e1, e2)}$ and 
$R_{\mathrm{E,cone}}^{\mathrm{max}(e1, e2)}$ are
the minimum and maximum value of $D_{\mathrm{ele}}$ and $R_{\mathrm{E,cone}}$
respectively, of the two electron candidates, which form the electron pair.
$B1$, $B2$ and $B3$ are background enhanced regions and
$S$ denotes the electron signal enhanced region, which is defined by 
 tight electron identification and isolation cuts.  
Templates for the determination of the background fractions are obtained
from Monte Carlo
simulations, see figure~\ref{fig:LightQuarkSeparation}. 
More than $70\%$ of the beauty, charm and $\JPSI$ events populate the signal 
enhanced bin $S$, since these events contain genuine electrons.
The uds events are  enriched in the three background bins $B1$, $B2$ and $B3$,
due to misidentified electrons. The measured number of  events in these three 
background bins constrain mainly the uds background  fraction.

\renewcommand{\arraystretch}{1.0} 
\begin{table}[tt]
\begin{center}
\begin{tabular}{|r|c|c|}
\hline
\raisebox{-1.0ex}{$R_{\mathrm{E,cone}}^{\mathrm{max}(e1, e2)}$} 	& 
\multicolumn{2}{|c|}{\raisebox{-1.0ex}{$D_{\mathrm{ele}}^{\mathrm{min}(e1, e2)}$}} 	 \\
									& $0.825-0.875$	& $0.875-1.0$ \\
\hline 
$150 - 350 \%$ 			& $B1$ 				& $B3$ \\
$0 -150 \%$ 				& $B2$ 				& $S$\\
\hline

\end{tabular}
\caption{Definition of the four regions used to constrain mainly the uds background.}
\label{tab:udsTemplatesDef}
\end{center}
\end{table}

\subsubsection{Heavy Quark Fractions}

In the signal enhanced region $S$, the individual contributions 
from beauty, charm and $\JPSI$  can be
disentangled by investigating the charge product, $q_{e1} \cdot q_{e2}$, of the
$e^\pm$-candidates, their azimuthal separation 
$\Delta \phi_{e1,e2} = |\phi_{e1} - \phi_{e2}|$, and their invariant mass $m_{e1,e2}$. 
Templates of the different background sources 
and of the beauty signal, which are all restricted to the signal enhanced region $S$, are shown in 
figure~\ref{fig:FlavourSepPropaganda} as function
of the invariant mass  $m_{e1,e2}$ and the signed azimuthal 
separation $\Delta \phi_{e1,e2} \cdot q_{e1} \cdot q_{e2}$. 

The different templates show specific characteristics: $\JPSI$ events have oppositely
charged electrons and cluster at $m_{e1,e2} = m_{\JPSI}$, whereas background from
open charm production covers a large mass range. Electrons from open charm
decays are found mostly back-to-back and with opposite charge sign, 
whereas electron pairs from beauty decays populate all $\Delta \phi_{e1,e2}$ values with
both charge sign combinations. Both charge products are also found
in the uds background, which however populate on average regions with smaller
$m_{e1,e2}$ values. Large values of $m_{e1,e2}$ are solely populated by beauty decays.

These distinct signatures of the individual background sources, i.e.~uds, 
$J/\psi \rightarrow ee$ and $c\bar c \rightarrow ee$,  are exploited
by dividing the signal enhanced region $S$ into 12 subregions 
($S1$ to $S12$) 
as shown in figure~\ref{fig:FlavourSepPropaganda}.
%
In the following the three background enhanced bins $B1$-$B3$ and the 12 signal enhanced bins
$S1$-$S12$ are referred to as 
``Flavour Separator'', 
for which templates are derived.

\subsection{Unfolding}
\label{sec:unfolding}

Using an unfolding procedure the number of background events $N_{\mathrm{uds}}$,
$N_{\JPSI}$, $N_{\mathrm{charm}}$ and the number of beauty  events $N_{\mathrm{beauty}, i}$
in four bins of $\langle P_T(b) \rangle$ are derived.
A regularized unfolding procedure is used with a smoothness condition.
The procedure is explained in appendix~\ref{Appendix Unfolding}. 
All efficiency corrections and migration effects are described by the response matrix 
$\mathbf{A}$, which correlates 
the number of reconstructed events  
in the  Flavour Separator distribution in bins of $m_{T, \mathrm{rec}}$,
represented by
the vector $\mathbf{y}$, with
the  distribution $\mathbf{x}$ on parton level via the matrix equation
\begin{linenomath*}
\begin{equation}
\label{eq:matrixequation}
\mathbf{y}=\mathbf{Ax+b} \quad .
\end{equation}
\end{linenomath*}
The  vector $\mathbf{x}$, defined as
 $\mathbf{x}^T=(\mathbf{x}_{\mathrm{beauty}}^T, x_{\mathrm{charm}}, x_{\JPSI}, x_{\mathrm{uds}})$, 
contains contributions from beauty  binned
in $\langle P_T(b) \rangle$, charm, $\JPSI$ and uds.
The contribution from beauty ($\mathbf{x}_{\mathrm{beauty}}$)
is defined according to the phase space given in 
table~\ref{tab:PhaseSpace}.
The vector $\mathbf{b}$ contains the background contribution from DIS events, 
which is taken
from  simulation. All other background
contributions  are incorporated in the response matrix and are determined by unfolding.

\renewcommand{\arraystretch}{1.2} 
\begin{table}[tt]
\begin{center}
\begin{tabular}{|c|}
\hline
Phase Space  \\
\hline
$Q^2\leq1\gevsq$\\
$0.05\leq y \leq0.65$\\
$|\eta(b)|, |\eta(\bar{b})|\leq 2$\\
\hline
\end{tabular}
\caption{Definition of the kinematic range of this measurement.}
\label{tab:PhaseSpace}
\end{center}
\end{table}

Signal and background templates as function of nine $m_{T,\mathrm{rec}}$ bins
are generated by Monte Carlo simulations and fitted to the data.
The unfolding procedure uses in total 
$N_{m_{T, \mathrm{rec}}} \times N_{\mathrm{Flavour \ Separator}} = 9 \times 15$
input bins and determines the three background fractions  and the number of beauty events 
$N_{\mathrm{beauty}, i}$ in four $\langle P_T(b) \rangle$ bins.
A schematic representation of the procedure is shown in figure~\ref{fig:ResponseMatrix_Schematic}.
In this procedure the $m_{T, \mathrm{rec}}$ dependences of the
different background contributions from uds, $J/\psi \rightarrow ee$ and $c\bar c \rightarrow ee$
 are fixed by the Monte Carlo predictions. The latter is
motivated by recent measurements of the differential cross sections of charm
production at HERA, which were found to be consistent with theoretical models
and Monte Carlo programs used in this analysis \cite{Aaron:2010gz,Dstargp}.

\renewcommand{\arraystretch}{1.2} 
\begin{table}[tt]
\begin{center}
       \begin{tabular}{| r | r   r  r  r| r |}
       \hline
         \multicolumn{5}{|c|}{\bf \boldmath Background Correlations }  & \multicolumn{1}{|c|}{\bf \boldmath Fractions} \\
      \hline
      \hline
  						&beauty 	& charm 		& $J/\psi$	& uds & [$\%$]\\
  		\hline					
  		beauty				&	1		&	-0.46	&	-0.18	&	-0.18	&	$25.8 \pm 3.6$ 		\\
  		charm				&			&	1			&	-0.03	& -0.27 	&	$17.6 \pm 3.3$				\\
  		$J/\psi$			&			&				&	1			&	0.03 		&	$29.0 \pm 2.1$				\\
  		uds					&			&				&				&	1			&	$25.3 \pm 3.0$				\\
      \hline
    \end{tabular}
    \caption{Correlations between the signal (beauty) and the different background contributions,
     and the determined relative fractions with their errors for the  data sample. The fraction of
     DIS events (not given in the table)
     is $2.3\%$. }
    \label{tab:Correlations}
  \end{center}
\end{table}

The fitted beauty signal and background contributions are shown in 
figure~\ref{fig:BeautyTag_Schematic} in the three background 
and in the signal enhanced regions. 
The event numbers resulting from the fitted fractions show very good agreement 
with the data considering statistical errors only. 
A clear enhancement of the genuine electron signal due to the tightening of the
electron identification cuts is seen when going from the first
background enhanced region ($B1$), which contains more than 80\% uds background to the  
signal enhanced region ($S1$-$S12$) with less than 20\% of uds background.

The correlations between the beauty signal and
the background sources, which are largest between  beauty  and 
charm, are
given in table~\ref{tab:Correlations} together
with the determined fractions of the selected data sample.

The distribution of the data as a function of the Flavour Separator is shown in 
figure~\ref{fig:ControlHistoSelection_FlavourSeparator} 
together with the result from the fit of the beauty and the 
various background contributions.
Good agreement  is found 
considering statistical errors only.

Control distributions of electron 
variables are presented in figure~\ref{fig:ControlDistrEle}  for the electron 
enriched signal region ($S1$-$S12$).
The data are compared to the simulated beauty signal  and background distributions
using the quark flavour decomposition determined by the unfolding procedure.
The main characteristics of the signed variables 
$\Delta \phi_{e1,e2} \cdot q_{e1} \cdot q_{e2}$ and $m_{e1,e2} \cdot q_{e1} \cdot q_{e2}$,
and $P_T(e)$ and $\vartheta(e)$ 
are well described by the Monte Carlo simulation.
In figure~\ref{fig:ControlDistrTracks} additional control distributions are
presented showing
the $P_T$-spectra of the three highest \mbox{$P_T$-tracks}.
These distributions are strongly dependent on the  track trigger conditions used,
and  imperfections of the trigger simulation would be visible here.

Reasonable agreement between the data and the Monte Carlo simulation is
obtained in all  distributions which gives confidence that the Monte Carlo
simulation is able to correctly model the detector response used for the
unfolding procedure. 


\subsection{Cross Section Determination and Systematic Uncertainties}
\label{sec: Cross Section Determination and Systematic Uncertainties}

The visible cross section is measured for the phase space as defined in
table~\ref{tab:PhaseSpace}. 
The bin-averaged differential cross section is obtained as
\begin{linenomath*}
\begin{align}
\label{eq: Xsection1}
 \dfrac{\mathrm{d} \sigma(ep \rightarrow e\, b\bar{b} X)}{\mathrm{d}\langle P_T(b) \rangle} = 
 \dfrac{N_{\mathrm{beauty}, i}}{\mathcal{L}  \cdot BR \cdot \Delta \langle P_{T, i}(b) \rangle} \quad,
\end{align}
\end{linenomath*}
where $\mathcal{L}$ is the luminosity, $\Delta \langle P_{T, i}(b) \rangle$ the bin width, $N_{\mathrm{beauty}, i}$ 
the number of unfolded signal
events in the corresponding bin and $BR= 6.17\%$ the effective branching fraction computed from~\cite{Nakamura:2010zzi}
for a $b\bar{b}$ pair decaying into at least two electrons. For the calculation of cross section 
uncertainties correlations between bins are taken into account.

The systematic uncertainties related to the measurement of the number of
$b\bar{b}$ signal events are listed in
the following. The effect on $N_{\mathrm{beauty, i}}$ is calculated by varying the 
sources of uncertainties in the simulation and by propagating these variations 
to the measurement through the response matrix
\textbf{A} and the background term \textbf{b} in equation~\ref{eq:matrixequation}.
\begin{itemize}
 \item The uncertainty on the electron identification is determined using $\JPSI \rightarrow e^+e^-$ 
   events (see figure~\ref{fig:ElePiDiscriminator}), 
   by comparing the distributions of the electron discriminator $D_{\mathrm{ele}}$ between
   data and Monte Carlo around each of the used  cut values of $D_{\mathrm{ele}}=0.875$ 
   and $D_{\mathrm{ele}}=0.825$. 
   The cut on $D_{\mathrm{ele}}$ is varied in MC by $\pm 0.025$ which covers any possible 
   shift in the $D_{\mathrm{ele}}$ distribution between data and simulation. 
   This   cut variation on $D_{\mathrm{ele}}$ propagated to the total beauty cross section 
   results in an uncertainty of $\pm 6.8\%$.  
 \item The uncertainty on the track finding efficiency of electrons is 
   conservatively estimated to be $2\%$
   per track resulting in an uncertainty of the total beauty cross section of $\pm 4\%$.
 \item The trigger uncertainty of the FTT at levels L1 and L2 are about
   $1-2\%$ each.
   The dominating contribution to the trigger
   uncertainty is due to the uncertainty of the calibration constants of the
   JT used at L3.
   To quantify this uncertainty, the JT calibration constants used in the
   simulation are varied
   by scaling the default calibration constants by $15\%$~\cite{msauter}.
   The systematic error on the total beauty cross section due to the uncertainty on
   the trigger efficiency is determined to be $\pm 8.6\%$.
 \item 
   Model uncertainties of the beauty signal are determined by comparing the
  default response matrix computed by taking the average of the two Monte Carlo samples
   (CASCADE and PYTHIA) with two alternative response matrices based on one
   of the Monte Carlo samples.
   The relative maximum difference with respect to the default
   response matrix is computed for each entry of the matrix and propagated
   to a model uncertainty on the total beauty cross section of $\pm 3.3\%$.
 \item The uncertainty of the charm contribution is evaluated from the relative difference
between the Monte Carlo generators CASCADE and PYTHIA in a similar way as for
the beauty signal. The systematic error on the extracted 
total beauty cross section due to the charm model is determined to be $\pm 3.6\%$.
  \item The uncertainty due to the fragmentation function of the heavy quarks is
estimated by reweighting the events according to the longitudinal
string momentum fraction $z$ carried by the heavy hadron in the Lund
model using weights of $(1 \mp 0.7)\cdot(1-z) + z \cdot (1 \pm 0.7)$
for charm quarks and by $(1 \mp 0.5)\cdot(1-z) + z \cdot (1 \pm 0.5)$
for beauty quarks~\cite{:2009ut}. 
The corresponding systematic error 
on the total beauty cross section is determined to be $\pm 3.4\%$ resulting from charm and 
$\pm 2.2\%$ from beauty fragmentation uncertainty.
 \item The uncertainty on the contribution from the remaining uds background due 
 to misidentified and  real electrons, was determined by varying their relative contributions 
 by a factor two up and down. The corresponding systematic error on the total 
 beauty cross section is determined to be $\pm 3.4\%$.
 \item CASCADE does not fully simulate the radiative tail of $\JPSI \rightarrow ee$ events.
  To estimate the uncertainty on the modelling of it,
 weights are applied, which are obtained from an elastic $\JPSI\rightarrow ee$ simulation 
 with radiative  QED corrections~\cite{Barberio:1993qi}. The systematic uncertainty on 
 the total beauty cross section is estimated to be $\pm 3.5\%$.
\item The uncertainty of the DIS-background, represented in equation~\ref{eq:matrixequation} 
by the vector $\mathbf{b}$, is taken to be $100\%$ and
  results in an error on the total beauty cross section of $\pm 4.5\%$.
 \end{itemize}

In addition, a global normalisation uncertainty of $4.1\%$ is included with 
contributions from the integrated luminosity uncertainty  of $2.7\%$ and from the uncertainty
on the semi-electronic branching fractions of $3.0\%$.

Adding all above
contributions in quadrature gives a total  systematic error
of $15.4\%$ on the total beauty cross section.

\section{Results}

The differential cross section $\mathrm{d}\sigma(ep \rightarrow e\, b\bar{b} X)/\mathrm{d} \langle P_T(b) \rangle$ is measured 
in the phase space defined in table~\ref{tab:PhaseSpace} using the 
unfolding procedure as described in section~\ref{sec:unfolding}. 

The result  is shown in table~\ref{tab:CrossSection_4bins_reg} together with statistical and 
total errors and the coefficients describing the
statistical correlations between bins.
In order to cross check the unfolding procedure the cross section extraction is repeated
without regularisation condition. 
The results obtained with and without regularisation  are found to be consistent
within the uncertainties. 

The measured differential beauty cross section is compared in figure~\ref{fig:DiffXsectionPtB} 
with an NLO QCD prediction in the fixed flavour
number scheme as calculated by the program FMNR.
The figure also shows the ratio of the measured cross section and the 
NLO QCD cross section. 
The uncertainties of the measurement are smallest at low $\langle P_T(b) \rangle$,
where the cross section is largest. 
The theoretical prediction of the differential cross section agrees  with the 
measurement within the large experimental and theoretical uncertainties. 
The prediction has a tendency to be below the data, a trend also observed in
 previous beauty cross section measurements at large transverse beauty momenta.

By integrating the differential cross section
the total inclusive beauty photoproduction cross section is measured as:
\begin{linenomath*}
\begin{align}
\sigma(ep \rightarrow e\, b\bar{b} X) = 3.79 \pm 0.53 \, \mathrm{ (stat.)} \pm 0.58 \, \mathrm{ (sys.)} \nbarn \quad,
\end{align}
\end{linenomath*}
to be compared with the NLO prediction obtained from FMNR of
 $\sigma(ep \rightarrow e\, b\bar{b} X) = 2.40_{-0.49}^{+0.55}\nbarn$. 
The measured cross section is higher, but within the large experimental and theoretical
uncertainty consistent with the NLO expectation.

\section{Conclusions}
The inclusive and differential cross section of
beauty photoproduction was measured in the di-electron final state, 
using the H1 detector at the HERA collider. 
The cross section is measured  as function 
of the quadratically averaged transverse momentum of the produced beauty quarks 
$\langle P_T(b) \rangle$,
with a special focus on the low $\langle P_T(b) \rangle$ regime.
Background from uds, charm and $\JPSI$  production is determined exploiting angular,
charge and mass correlations of electron pairs in an unfolding procedure.

The measured cross section is compared to a  QCD prediction at NLO performed 
in the fixed flavour number scheme and evaluated with 
$\mu_{R} = \mu_{F} = 1/2 \, \sqrt{m_b^2 + \langle P_T(b) \rangle  ^ 2}$
as choice for the renormalisation and factorisation scale.
The NLO prediction lies below the data but within
the large experimental and theoretical  uncertainty  
they agree.

This measurement is in good agreement with previous measurements of beauty
photoproduction at HERA and it extends the previously experimentally accessible 
phase space towards 
the beauty production threshold.

\section*{Acknowledgments}
We are grateful to the HERA machine group whose outstanding
efforts have made this experiment possible. 
We thank the engineers and technicians for their work in constructing and
maintaining the H1 detector, our funding agencies for 
financial support, the
DESY technical staff for continual assistance
and the DESY directorate for support and for the
hospitality which they extend to the non-DESY 
members of the collaboration.

\newpage
\appendix
\section{Electron Discriminator Combination}
\label{Appendix Electron Discriminator Combination}
The track seeded and calorimeter based electron discriminator $D_{\mathrm{calo}}$ and
the electron discriminator $D_{\mathrm{d}E/\mathrm{d}x}$, based on the specific energy loss
measurement in the CTD, are mapped
to a combined discriminator using the expression
\begin{linenomath*}
\begin{align}
D_{\mathrm{ele}}(a, b, c, d) = \frac{|(1-D_{\mathrm{calo}})^a -1 |^c \cdot |(1-D_{\mathrm{d}E/\mathrm{d}x})^b -1 |^d}
{((1-D_{\mathrm{calo}})^a-1) \cdot ((1-D_{\mathrm{d}E/\mathrm{d}x})^b-1) +  (1-D_{\mathrm{calo}})^a  \cdot (1-D_{\mathrm{d}E/\mathrm{d}x})^b }
 \label{eq: DiscrComb}
\  \mathrm{,}
\end{align}
\end{linenomath*}
which for the parameter choice  $a=b=c=d=1$ corresponds
to Bayes' theorem:
\begin{linenomath*}
\begin{align}
D_{\mathrm{ele}}(1, 1, 1, 1) = \frac{|D_{\mathrm{calo}}| \cdot |D_{\mathrm{d}E/\mathrm{d}x}|}
{D_{\mathrm{calo}} \cdot D_{\mathrm{d}E/\mathrm{d}x} +  (1-D_{\mathrm{calo}}) \cdot  (1-D_{\mathrm{d}E/\mathrm{d}x}) }
 \label{eq: DiscrComb2}
\  \mathrm{.}
\end{align}
\end{linenomath*}
However, in order to obtain a sensible mapping behaviour of
$D_{\mathrm{calo}}$ and $D_{\mathrm{d}E/\mathrm{d}x}$  onto
$D_{\mathrm{ele}}$ when their respective values are close to 1 and 0 or
both of them are close to 1, the parameters $a=b=0.6$ and
$c=d=1.05$ are chosen.

\section{Unfolding procedure}
\label{Appendix Unfolding}
The differential cross section of beauty photoproduction is 
extracted from the measured di-electron spectrum 
using an unfolding procedure as implemented in TUnfold~\cite{TUnfold}. 

The vector $\mathbf{y}$, representing the number of measured events, 
is related via the matrix equation 
$\mathbf{y}=\mathbf{A} \cdot \mathbf{x} +\mathbf{b}$ to the true distribution
represented by a vector $\mathbf{x}$, which is determined by unfolding.
The response matrix $\mathbf{A}$ describes the detector acceptance,
contains all selection efficiencies and takes migration effects
between bins into account.
Additional background, not determined by the unfolding procedure and 
taken from external information, is represented by the vector $\mathbf{b}$.

An estimator $\mathbf{\hat{x}}$ of the true distribution $\mathbf{x}$ is obtained 
by unfolding the measured distribution~$\mathbf{y}$.
For the construction of $\mathbf{\hat{x}}$ additional assumptions, e.g. 
on the smoothness of the de-convoluted distribution (regularisation), and an additional constraint
on the number of observed events are applied.
In general $\mathbf{\hat{x}}$ is obtained by minimising a $\chi^2$ function given by:
\begin{linenomath*}
\begin{align}
\chi^2(\mathbf{\hat{x}}, \tau, \mu) &:= \chi^2_A(\mathbf{\hat{x}}) + \tau
\cdot \chi^2_L(\mathbf{\hat{x}}) + \mu \cdot
\chi^2_N(\mathbf{\hat{x}})        \label{eq:Unfolding6} \quad .
\end{align}
\end{linenomath*}
This equation describes the minimisation of the unfolding problem
$\chi^2_A(\mathbf{\hat{x}})$ with the two side conditions given by
$\chi^2_L(\mathbf{\hat{x}})$ and $\chi^2_N(\mathbf{\hat{x}})$.

The actual minimisation problem is defined by the standard $\chi^2$ function:
\begin{linenomath*}
\begin{align}
\chi^2_A(\mathbf{\hat{x}}) &:= \frac{1}{2} \left(\mathbf{y-b} -\mathbf{A
  \hat{x}}\right)^T \mathbf{V}^{-1} \left(\mathbf{y-b} -\mathbf{A
  \hat{x}}\right) \label{eq:Unfolding7} \quad 
\end{align}
\end{linenomath*}
with $\mathbf{V}=cov(y_i, y_j)$ being the covariance matrix of the data. 
This function minimises the deviation of the estimator $\mathbf{A \hat{x}}$ from
the measured, and background subtracted vector $\mathbf{y-b}$.

The additional  constraints are given by:
\begin{linenomath*}
\begin{align}
\chi^2_L(\mathbf{\hat{x}}) &:= \mathbf{\hat{x}}^T \mathbf{L} \mathbf{\hat{x}}  \label{eq:Unfolding8}\\
\chi^2_N(\mathbf{\hat{x}}) &:= \left( n_{\mathrm{obs}} - \sum_{j=1}^{m}  \left(\mathbf{A\hat{x}}\right)_j \right)^2  \label{eq:Unfolding9}
\mathrm{,}
\end{align}
\end{linenomath*}
with $\mathbf{L}$ being the regularisation matrix, $m$ the number of reconstructed bins 
and $n_{\mathrm{obs}}$ the total number of
observed events after background subtraction, which ensures that the total number of events is conserved.
Both functions enter equation~\ref{eq:Unfolding6} with the parameters
$\tau$ and $\mu$, where $\tau$ is often  denoted as regularisation parameter and 
$\mu$ as Lagrange Multiplier.

The $\chi^2_L(\mathbf{\hat{x}})$ function is a measure for the smoothness of the result. 
The matrix $L$ is chosen such that  the second derivative  of  
$\mathbf{\hat{x}}$ between bins describing beauty production is minimised. 
The regularisation parameter $\tau$ determines the strength of the smoothness constraint.
For the regularised unfolding $\tau$ is chosen such that the correlations of the covariance matrix of the
unfolded distribution $\mathbf{\hat{x}}$ are minimised~\cite{Unfolding_Blobel}.



\begin{sidewaystable}[!pht]
\begin{center}
 
       \begin{tabular}{|r| r @{, } r | r | r r  r |c||r|r|r|r|r|r|r|r|r|}
       \hline
         \multicolumn{17}{|c|}{\bf \boldmath H1 Beauty Photoproduction Cross Sections }  \\
      \hline
      \hline
    & \multicolumn{2}{|c|}{$\langle P_T(b) \rangle$} & $\langle P_{T, bc}(b) \rangle$&$\mathrm{d} \sigma / \mathrm{d} \langle P_T(b) \rangle$ & stat. & tot.  &  stat. corr. &
    $\delta_{sys.}^{b}$&$\delta_{sys.}^{c}$&
    $\delta_{sys.}^{uds}$&$\delta_{sys.}^{e-Id.}$&$\delta_{sys.}^{trig}$&
    $\delta_{sys.}^{fr. b}$&$\delta_{sys.}^{fr. c}$&$\delta_{sys.}^{\JPSI}$ &$\delta_{sys.}^{DIS}$\\
    & \multicolumn{2}{|c|}{[GeV]} & \multicolumn{1}{|c|}{[GeV]} &\multicolumn{3}{c|}{[pb/GeV]}                                            &           &
      [\%] &  [\%] &  [\%] &  [\%] &  [\%] &  [\%] &  [\%] &  [\%] &  [\%]\\
      \hline

 1 & $ [ 0.0 $&$  4.65 ] $     & $2.1$   &$ 487 $ & $ \pm  94 $ & $ \pm 123 $  &
 \begin{scriptsize}
  \begin{tabular}{ r @{=}r } 
  $\rho_{1,2}$& $0.02$ \\
  $\rho_{1,3}$& $-0.05$ \\
  $\rho_{1,4}$& $0.14$ \\
   \end{tabular} 
  \end{scriptsize}  
& $3$ & $-1$ & $0$ & $4$ & $11$ & $-2$ &  $4$ & $-5$ & $4$\\
\hline
                                           
 2 & $ [ 4.65 $&$  7.7 ] $     & $6.1$   &$ 358 $ & $ \pm  97 $ & $ \pm 112 $  & 
  \begin{scriptsize}
  \begin{tabular}{ r @{=}r } 
  $\rho_{2,3}$& $-0.38$ \\
  $\rho_{2,4}$& $0.25$ \\
   \end{tabular} 
  \end{scriptsize}  
 &  $4$ &  $7$ & $-1$ & $6$ & $5$ & $-2$ & $5$ & $-2$ &$4$\\
 
\hline

 3 & $ [ 7.7 $&$ 11.3] $    & $9.2$ &$ 92   $ & $ \pm  49 $ & $ \pm 65$ & 
   \begin{scriptsize}
  \begin{tabular}{ r @{=}r } 
  $\rho_{3,4}$& $-0.43$ \\
   \end{tabular} 
  \end{scriptsize} 
 & $-3$ & $15$ & $-34$& $21$ & $9$ & $-3$ & $-2$ & $0$&$10$\\
\hline

 4 & $ [ 11.3 $&$ 30.0] $ & $16.5$ &$ 5.9   $ & $ \pm  4.3 $ & $ \pm 5.3$ & 
   \begin{scriptsize}
  \begin{tabular}{ r @{=}r } 
   \end{tabular} 
  \end{scriptsize} 
 & $19$ & $36$ & $-2$& $20$ & $-15$ & $-1$ & $-12$ & $-2$&$1$\\
\hline

    \end{tabular}
    \caption{
Differential cross sections for the phase space
      defined in table~\ref{tab:PhaseSpace}  obtained from unfolding with
      regularisation condition. Also given are the statistical and
      total errors, the coefficients of the statistical
      correlations, the used bin boundaries in $\langle P_T(b) \rangle$ and the 
      corresponding bin centres~\cite{Lafferty:1994cj} $\langle P_{T, bc}(b) \rangle$. 
      The remaining columns list the 
    the	bin-to-bin correlated systematic uncertainties in the cross section measurement due to
uncertainties of the beauty ($\delta_{sys.}^{b}$) and charm
($\delta_{sys.}^{c}$) modelling, the uds background ($\delta_{sys.}^{uds}$), the electron
identification ($\delta_{sys.}^{e-Id.}$), the beauty ($\delta_{sys.}^{fr. b}$) 
and charm ($\delta_{sys.}^{fr. c}$) fragmentation, the modelling of the radiative tail
of $\JPSI \rightarrow e^+e^-$ events ($\delta_{sys.}^{\JPSI}$) and the 
DIS background ($\delta_{sys.}^{DIS}$).       
Not listed in the table is the  $4.1\%$ normalisation uncertainty.}
    \label{tab:CrossSection_4bins_reg}
  \end{center}
\end{sidewaystable}
\clearpage

\begin{figure}[htp]
\center
\epsfig{file=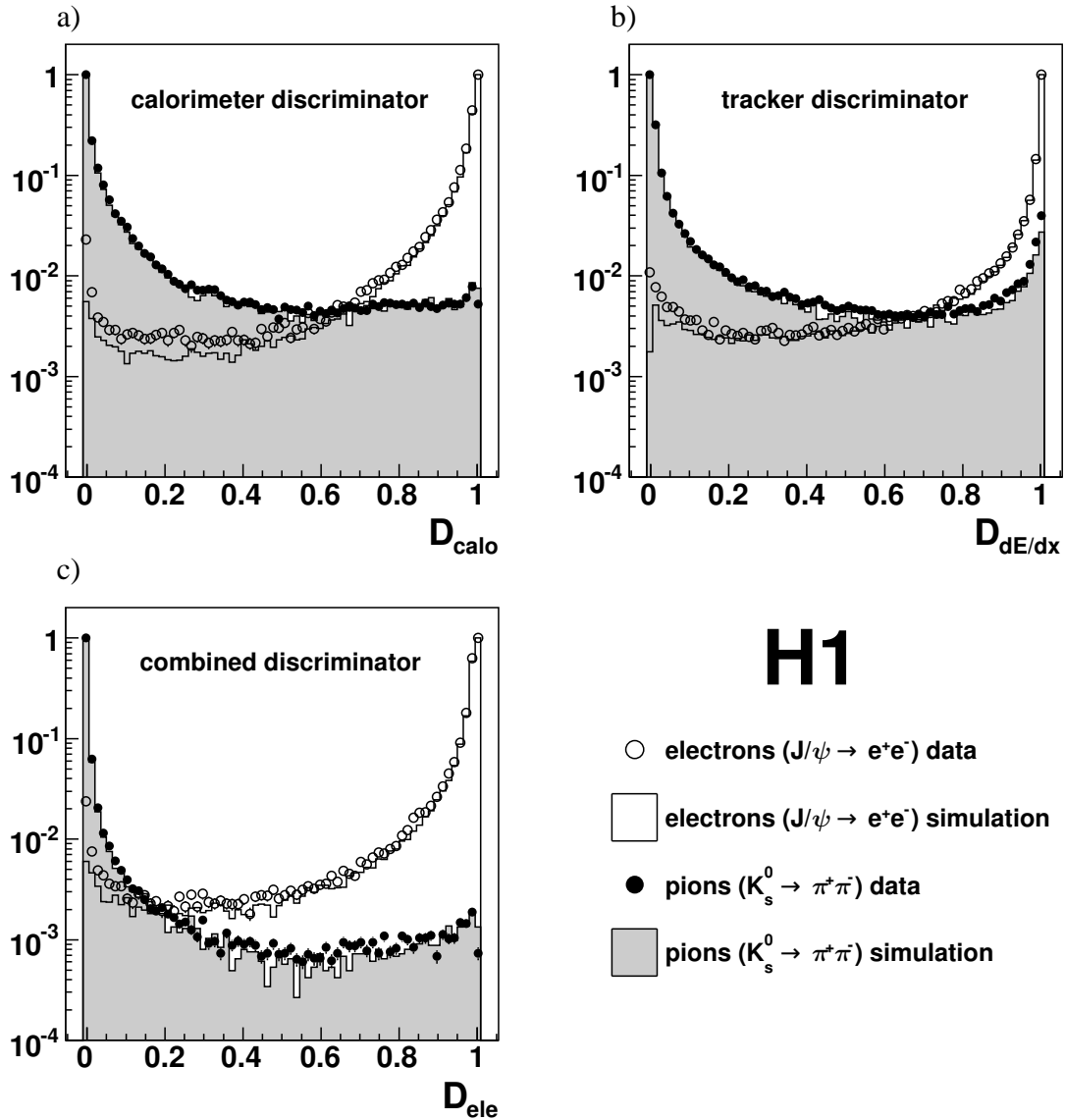 ,width=15cm}
\put(-140,150) {a)}
\put(-65,150) {b)}
\put(-140,75) {c)}
\setlength{\unitlength}{1cm}
\caption{Normalized discriminator distributions for the separation of electrons and pions
  as obtained from $\JPSI \rightarrow e^+e^-$ and 
  $K_s^0 \rightarrow \pi^+ \pi^-$ decays using the tag and probe method.
 a) the track seeded, calorimeter based discriminator $D_{\mathrm{calo}}$,
 b) the 
 discriminator $D_{\mathrm{d}E/\mathrm{d}x}$ based on the measurement of the specific energy loss in the CTD and
  c) their combination $D_{\mathrm{ele}}$.  
  Data are represented by circles and Monte Carlo simulations by
  histograms. 
}
\label{fig:ElePiDiscriminator} 
\end{figure}
\begin{figure}[htp]
\center
\epsfig{file=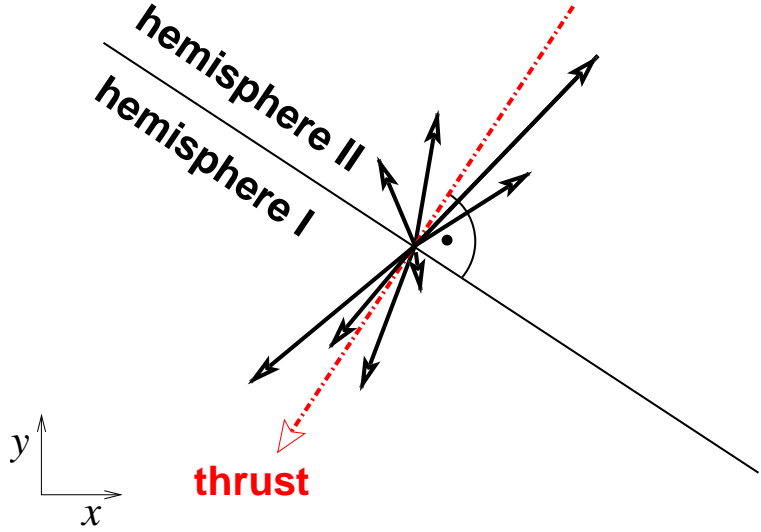 ,width=10cm}
\setlength{\unitlength}{1cm}
\caption{Schematic illustration of the determination of the thrust axis in the plane transverse to the $ep$~beams. 
The transverse thrust axis, indicated by the dashed arrow,
maximizes the sum of momenta projected onto it in this plane.
The thrust axis allows the event to be divided into two hemispheres, each containing the decay products of a beauty quark,
used to reconstruct the average transverse beauty mass $m_{T, \mathrm{rec}}(b)$ as defined in 
equation~\ref{eq: M_T_est}. 
}
\label{Fig:ThrustAxisSchematic}
\end{figure}
\begin{figure}[htp]
\center
\epsfig{file=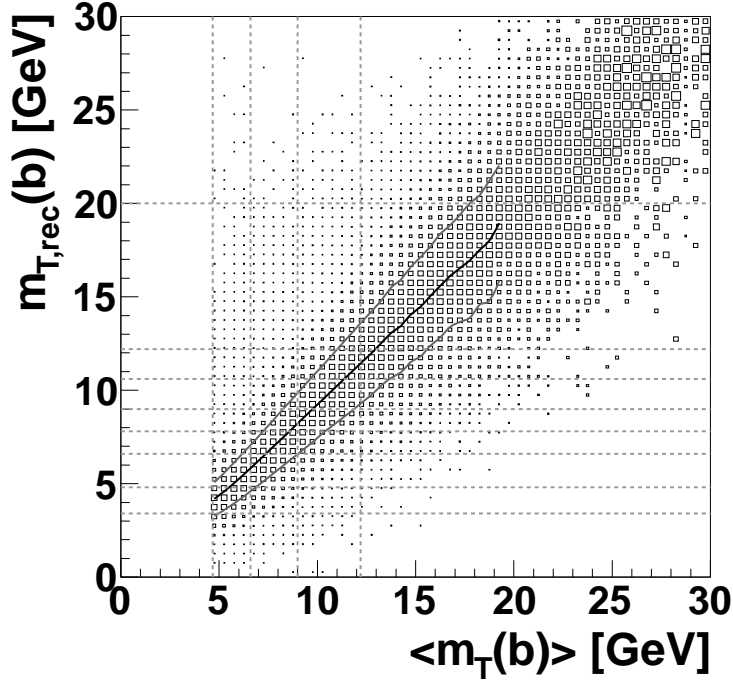 , width=10cm}
\setlength{\unitlength}{1cm}
\caption{Correlation between the reconstructed transverse beauty mass
  $m_{T, \mathrm{rec}}(b)$ and the transverse mass $\langle m_T(b) \rangle$ calculated from the quadratically averaged
  transverse momentum of the generated beauty quarks.
The  inner line on the diagonal indicates the correlation of $m_{T, \mathrm{rec}}(b)$ 
and $\langle m_T(b) \rangle$, and the outer two lines show the $1\sigma$ error band.
The used binning  (dotted grey lines) for the vectors $\mathbf{x}$ and $\mathbf{y}$ 
entering the unfolding procedure are also shown.}
\label{Fig:ThrustAxis}
\end{figure}
\begin{figure}[htp]
\center
\epsfig{file=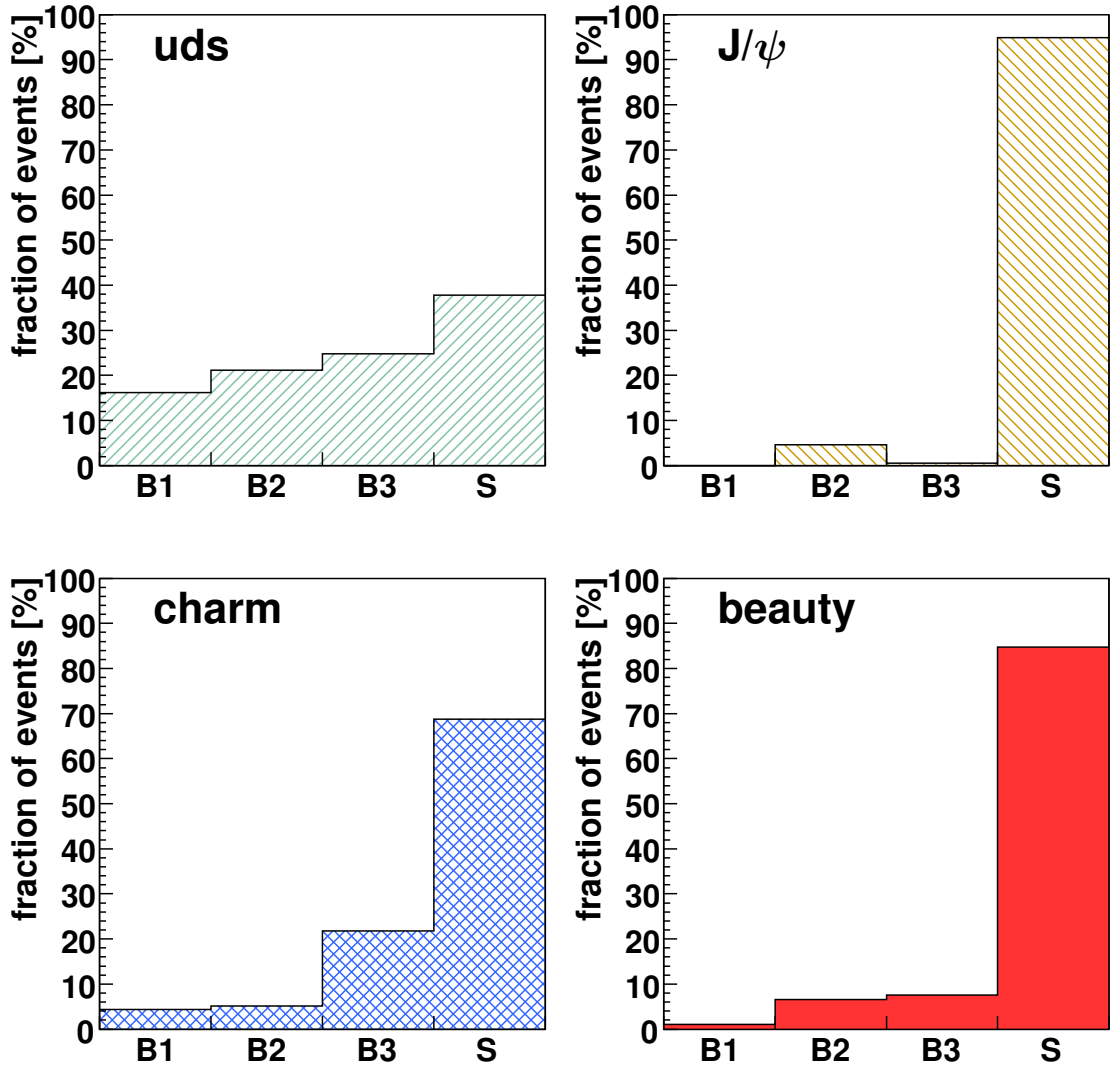 ,width=15cm}
\setlength{\unitlength}{1cm}
\caption{Templates used to separate the light quarks (uds) from the heavy quark flavours as obtained by the Monte Carlo simulation. For the definition of the background enhanced regions $B1$-$B3$ and 
the signal enhanced region $S$ see table~\ref{tab:udsTemplatesDef} and  text.
}
\label{fig:LightQuarkSeparation} 
\end{figure}
\begin{figure}[htp]
\center
\epsfig{file=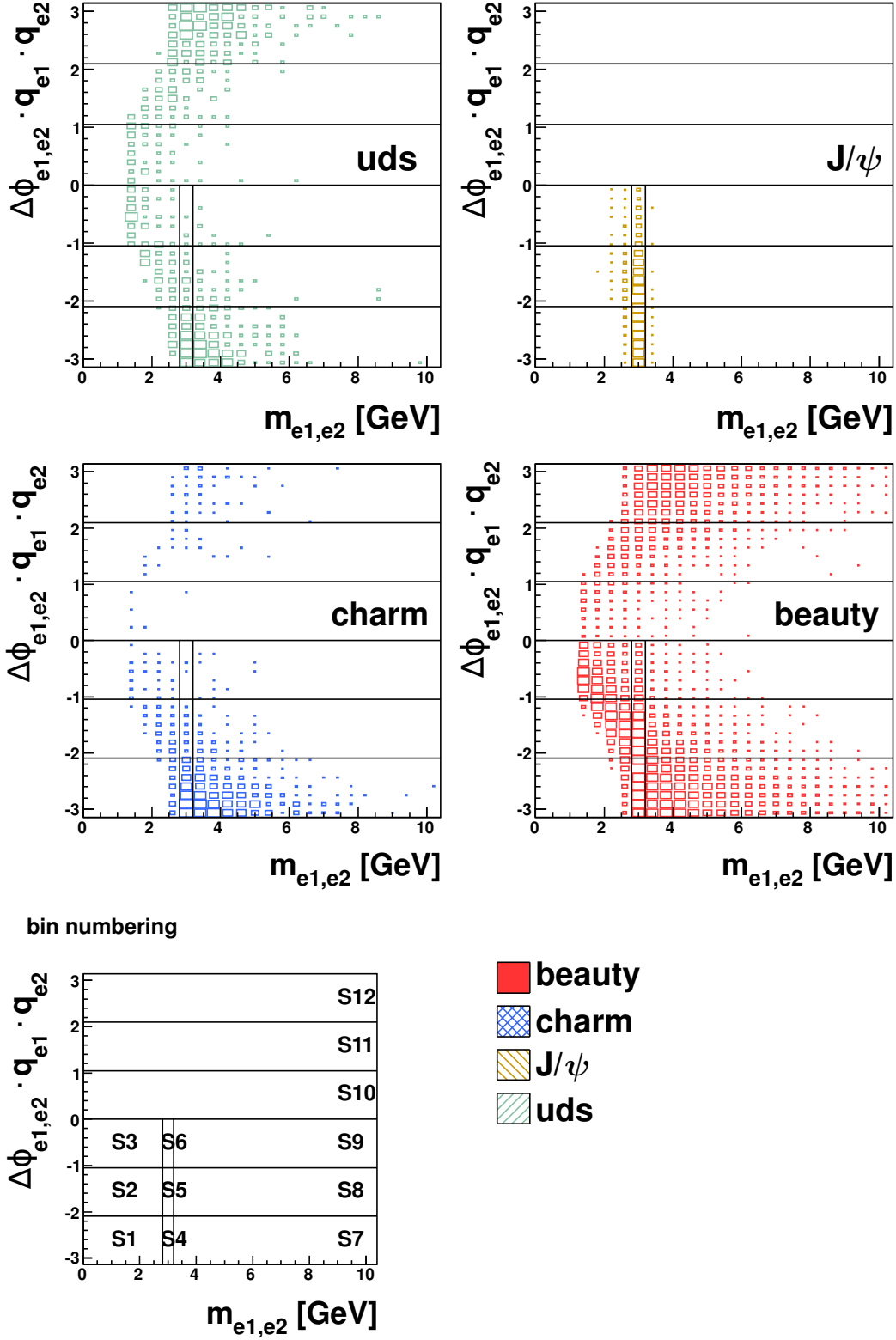 , height=21cm}
\setlength{\unitlength}{1cm}
\caption{Templates in the plane $\Delta \phi_{e1,e2} \cdot q_{e1} \cdot q_{e2}$, $m_{e1,e2}$
and restricted to the signal enhanced region $S$
used to separate the heavy quark flavours as obtained by the Monte Carlo simulation. 
Also shown is the bin numbering $S1$-$S12$
of the 12 subregions of $S$. For the definition of
the signal enhanced region $S$ see table~\ref{tab:udsTemplatesDef} and  text.
The two vertical lines indicate the peak invariant mass region of the
$\JPSI \rightarrow e^+e^-$ decays.
}
\label{fig:FlavourSepPropaganda} 
\end{figure}

\begin{figure}[htp]
\center
\epsfig{file=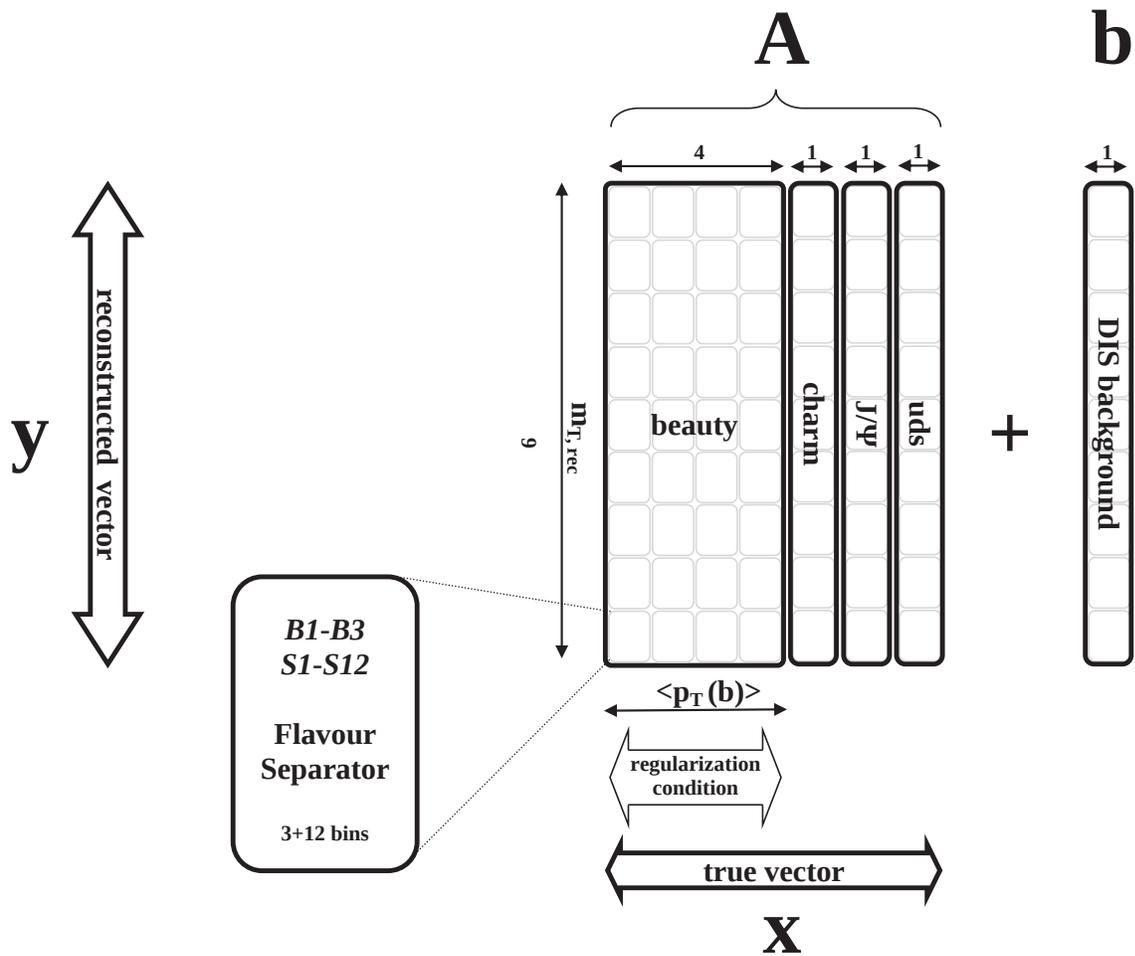 ,width=15cm}
\setlength{\unitlength}{1cm}
\caption{Structure of the response matrix $\mathbf{A}$ used to solve the matrix equation
$\mathbf{y}=\mathbf{A} \cdot \mathbf{x} +\mathbf{b}$ by unfolding.  The indicated numbers
 specify the number of used bins. The sub-binning  in $\mathbf{y}$ given by the 
 Flavour Separator allows the discrimination of the beauty signal from the uds, charm and $\JPSI$ backgrounds.
 See text for  details.
}
\label{fig:ResponseMatrix_Schematic} 
\end{figure}
\begin{figure}[!pht]
\center
\epsfig{file=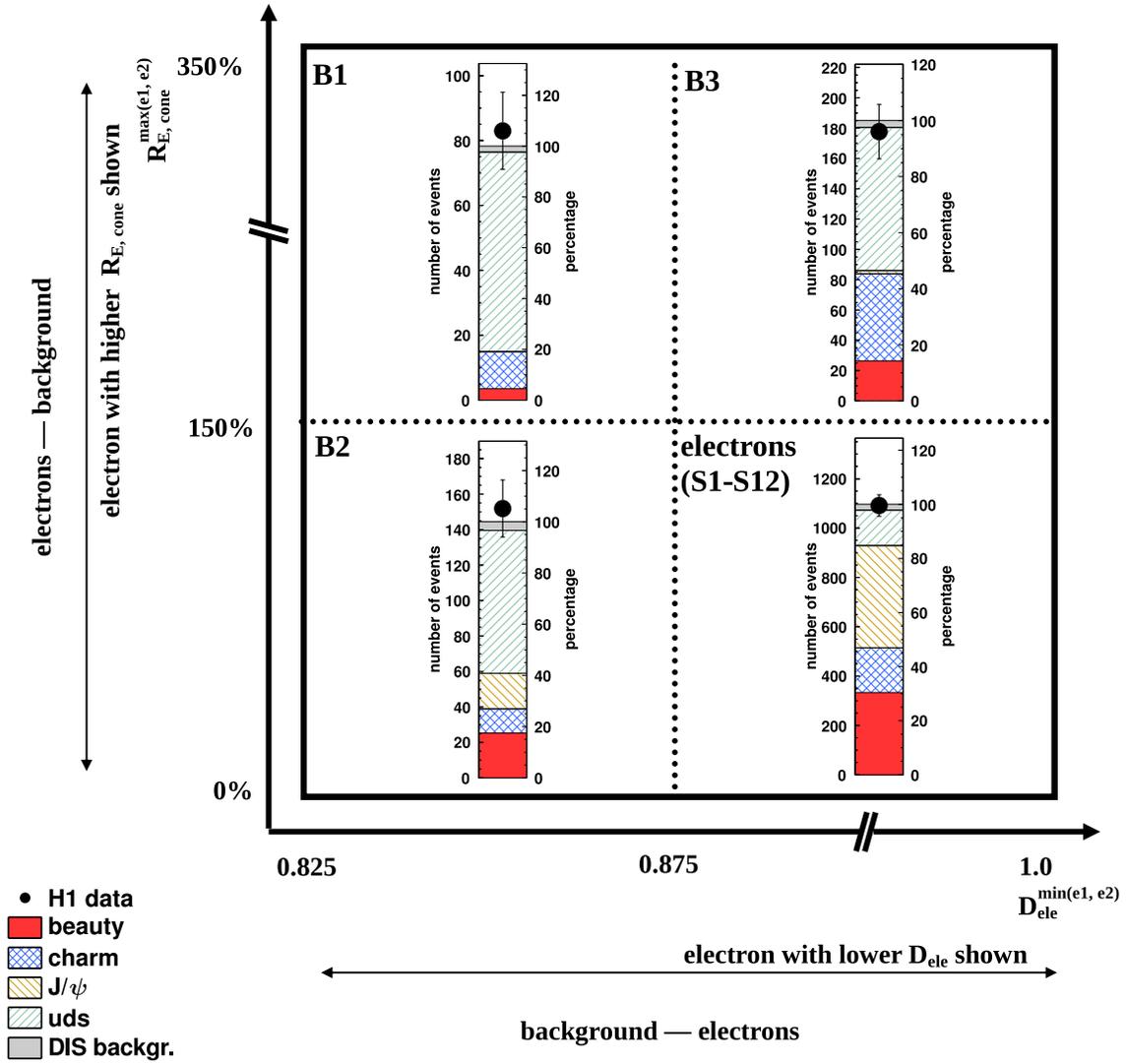 , width=15cm}
\setlength{\unitlength}{1cm}
\caption{Number of di-electron events in the background and signal enhanced regions 
as defined in table~\ref{tab:udsTemplatesDef}. Data are represented as points with the
statistical error indicated by the error bars.
Also shown in colour is the decompositions of the event yields as determined
by the unfolding procedure. }
\label{fig:BeautyTag_Schematic} 
\end{figure}

\begin{figure}[htp]
\center
\epsfig{file=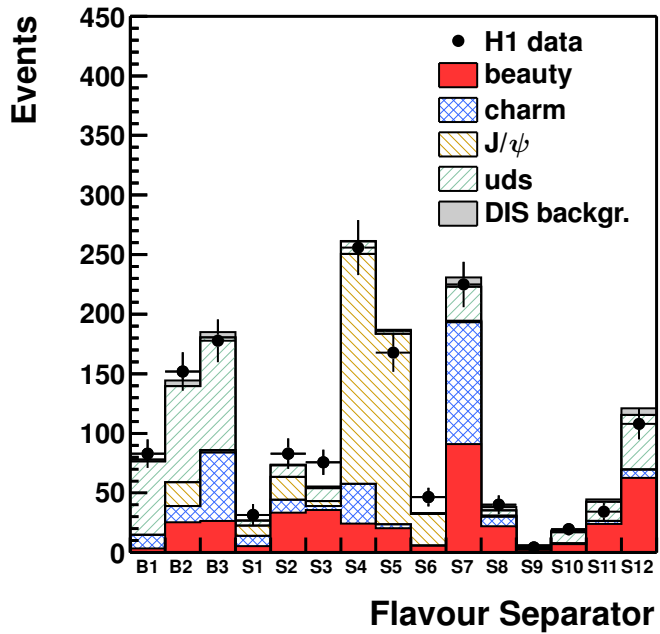 ,width=10cm}
\setlength{\unitlength}{1cm}
\caption{Number of di-electron events in the flavour separator histogram compared to
the number of fitted events and their decomposition.
Data are represented as points with the
statistical uncertainties indicated by the error bars.
The bin numbering scheme as defined in
figure~\ref{fig:FlavourSepPropaganda} and table~\ref{tab:udsTemplatesDef} is used. 
}
\label{fig:ControlHistoSelection_FlavourSeparator} 
\end{figure}

\begin{figure}[htp]
\center
\setlength{\unitlength}{7.5cm}
\epsfig{file=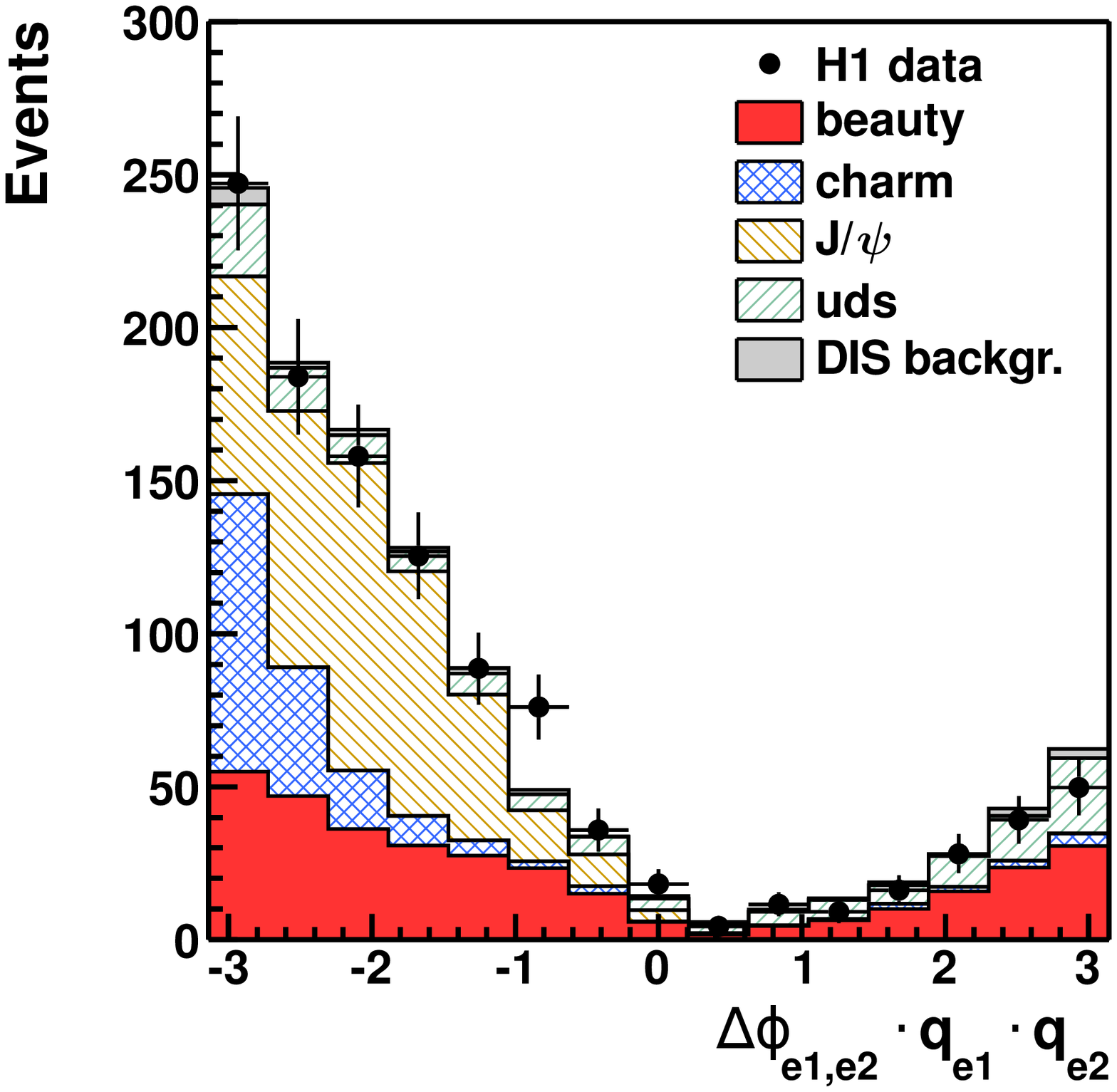,width=7.5cm}\put(-0.8,0.9) {a)}
\epsfig{file=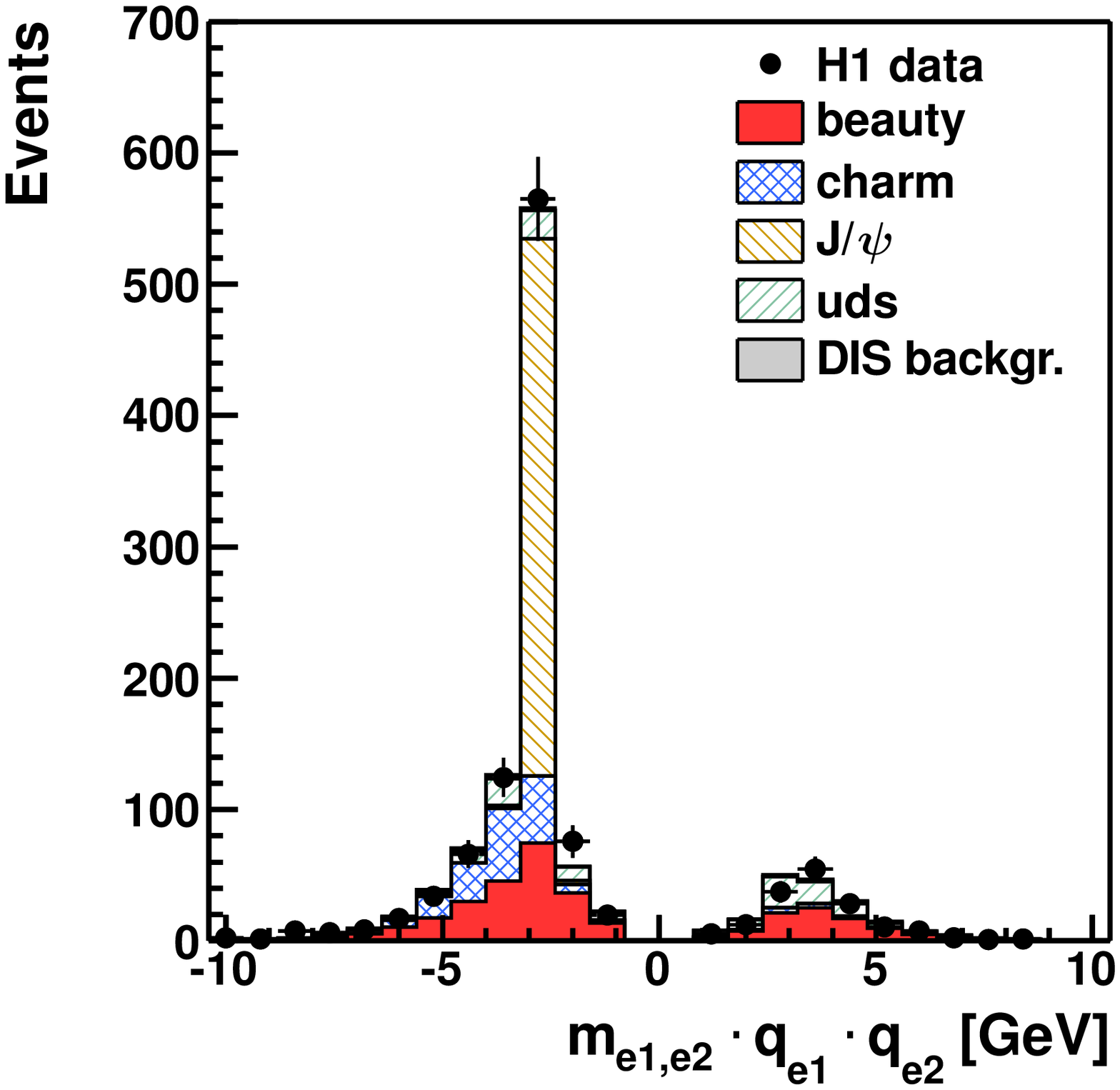,width=7.5cm}\put(-0.8,0.9) {b)}

\epsfig{file=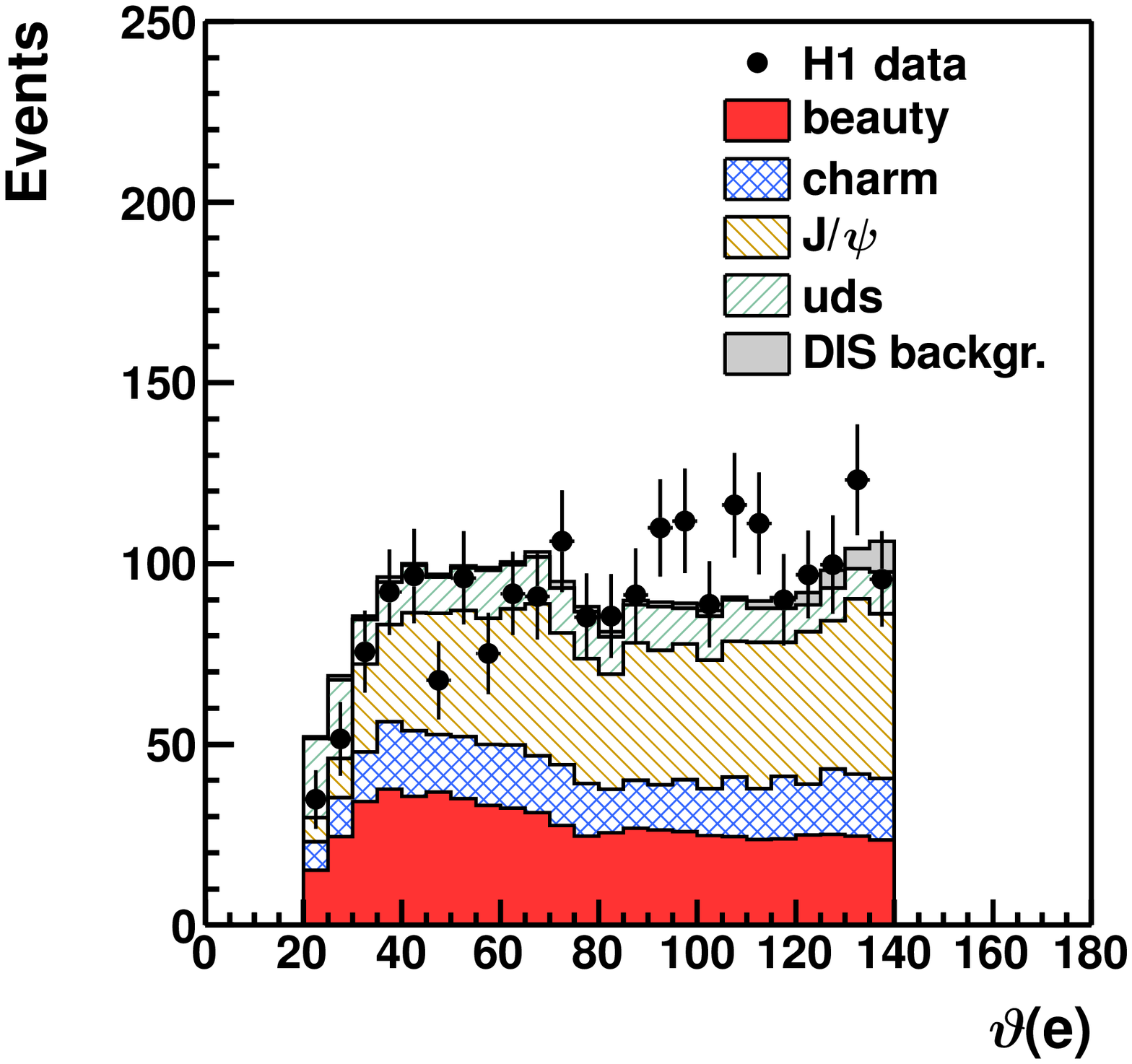,width=7.5cm}\put(-0.8,0.9) {c)}
\epsfig{file=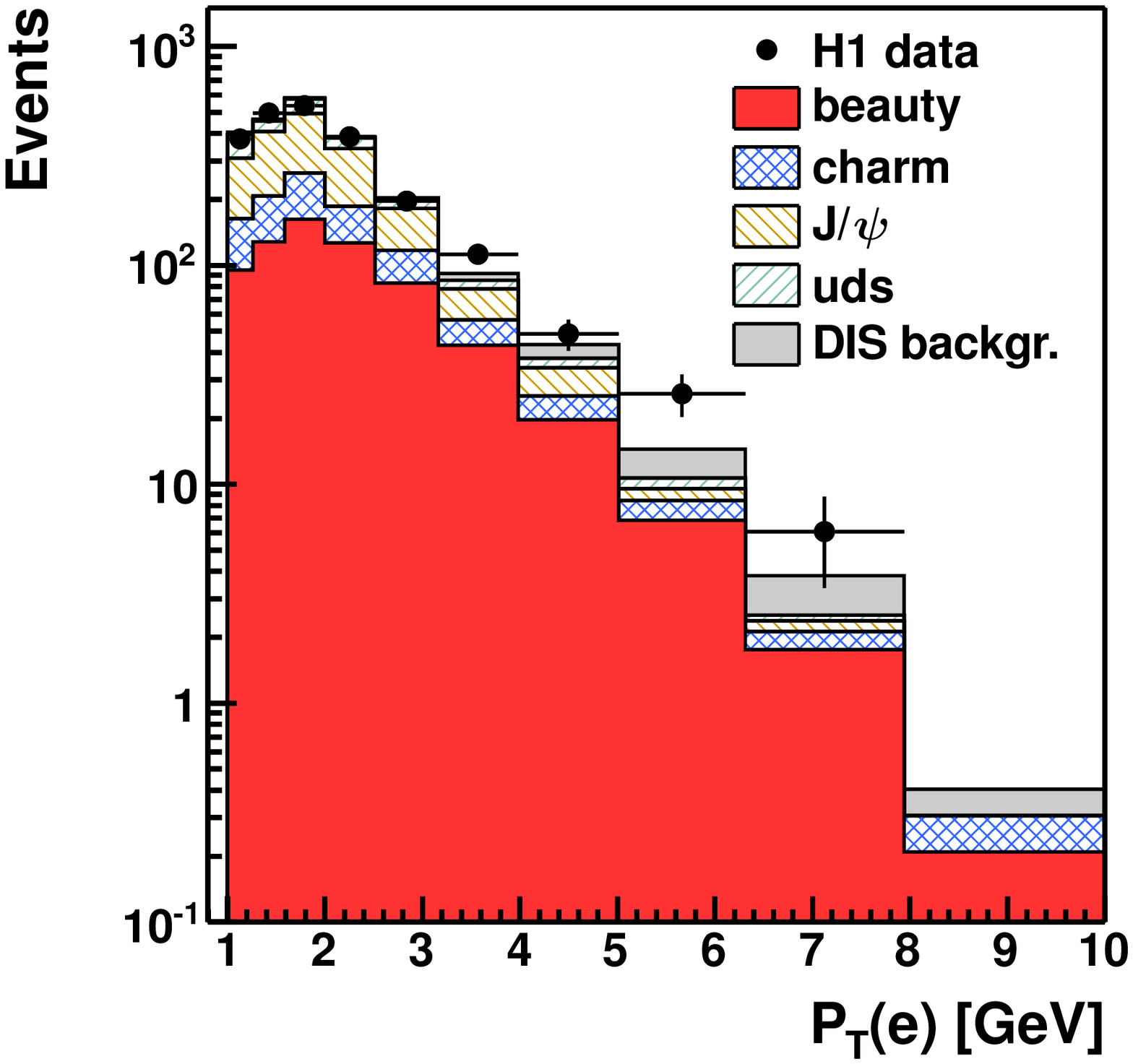,width=7.5cm}\put(-0.8,0.9) {d)}
\caption{
Control distributions of the electron candidates 
compared to 
Monte Carlo simulations using the quark flavour
decomposition determined by the unfolding procedure:
a) signed azimuthal separation  
$\Delta \phi_{e1,e2} \cdot q_{e1} \cdot q_{e2}$ defined by the charges multiplied with the azimuthal angle
difference of the two electron candidates, 
b) signed invariant mass $m_{e1,e2} \cdot q_{e1} \cdot q_{e2}$ defined by the charges multiplied with the
invariant mass of the two electron candidates, 
c) polar angle of the electron candidates and d) transverse momentum of the
electron candidates.
Data are represented as points with the
statistical uncertainties indicated by the error bars.
The distributions are restricted to 
the electron enriched region ($S$).}
\label{fig:ControlDistrEle} 
\end{figure}

\begin{figure}[htp]
\center
\setlength{\unitlength}{7.5cm}
\epsfig{file=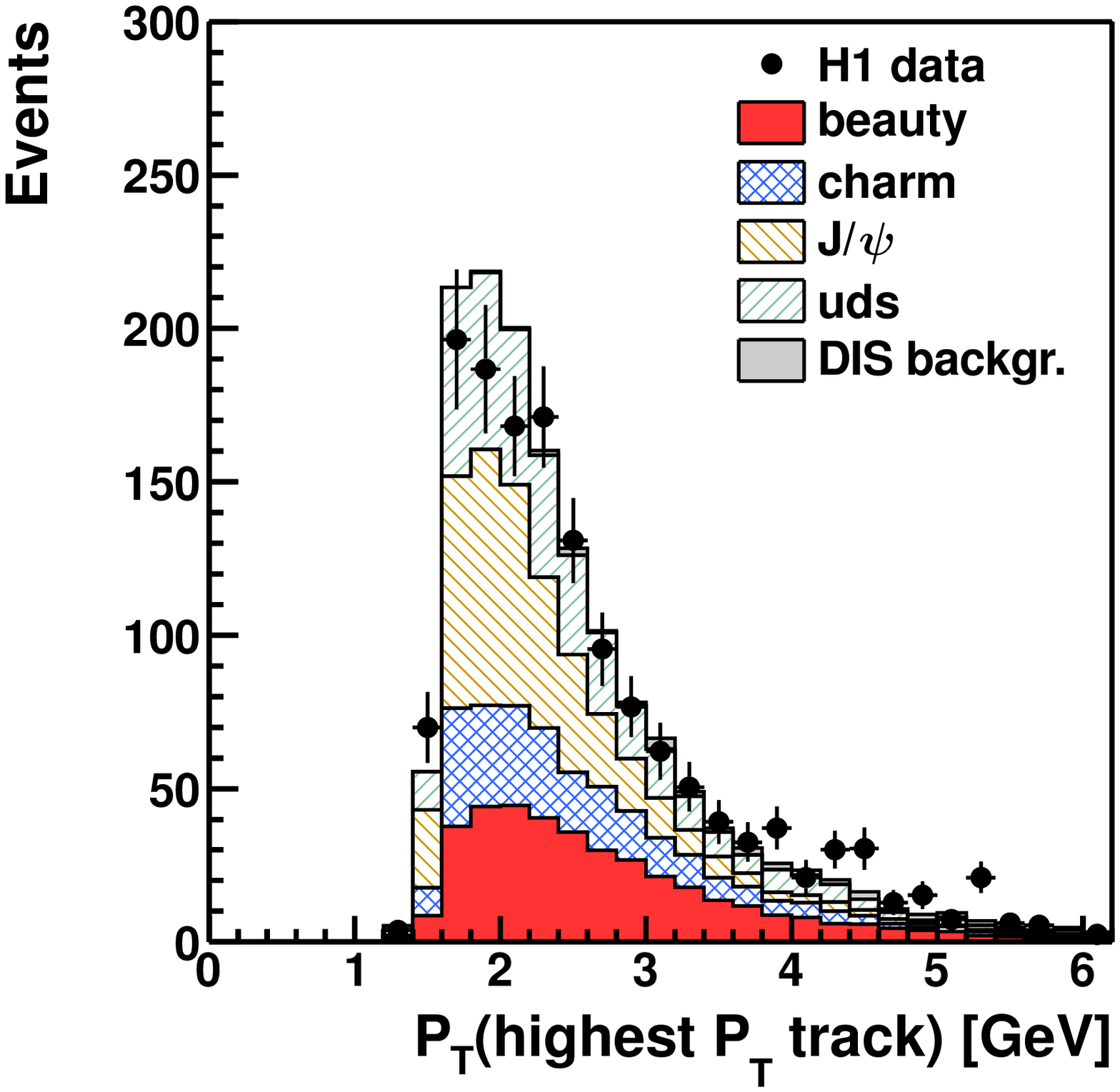,width=7.5cm} \put(-0.8,0.9) {a)}
\epsfig{file=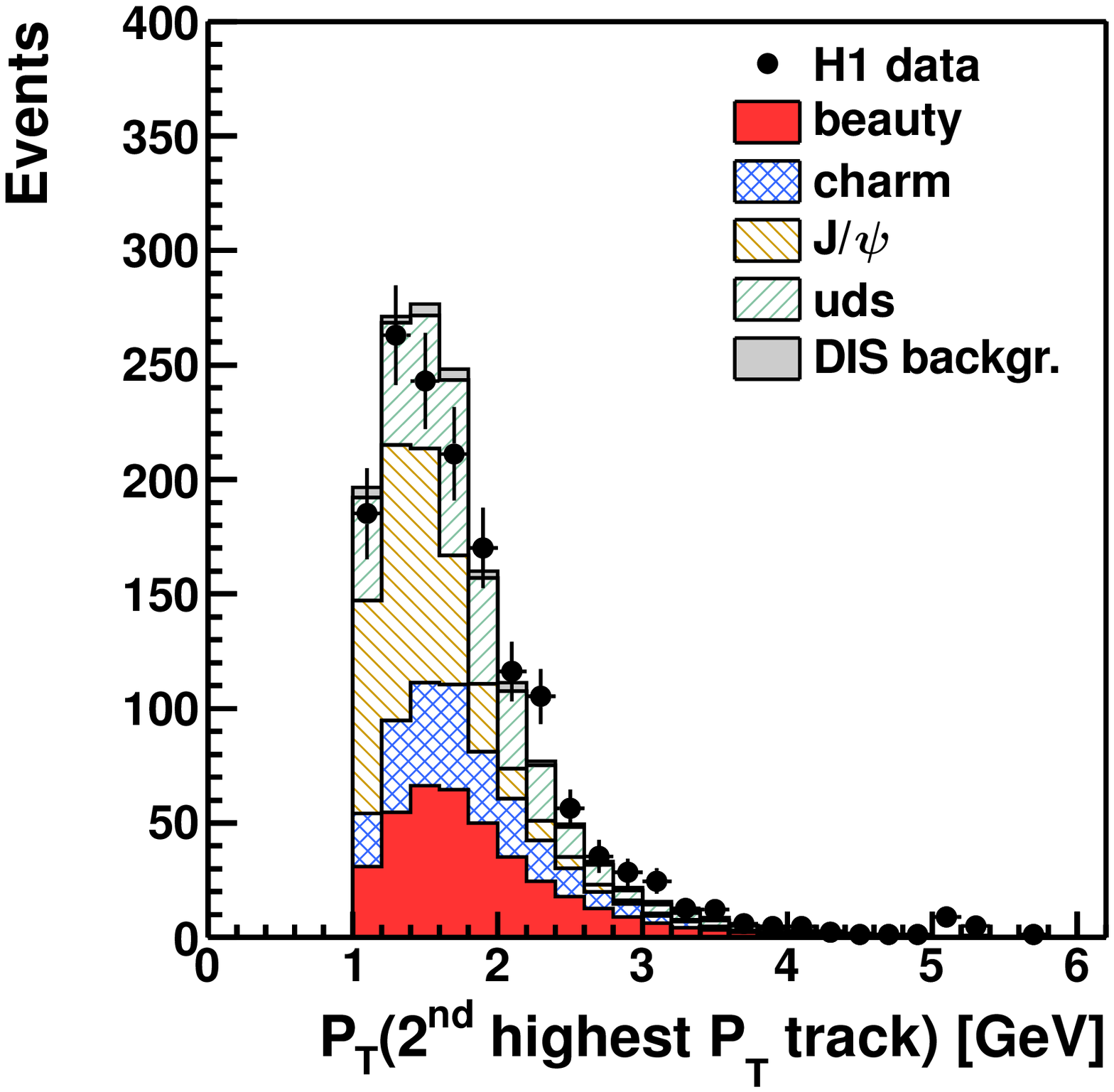,width=7.5cm}\put(-0.8,0.9) {b)}

\epsfig{file=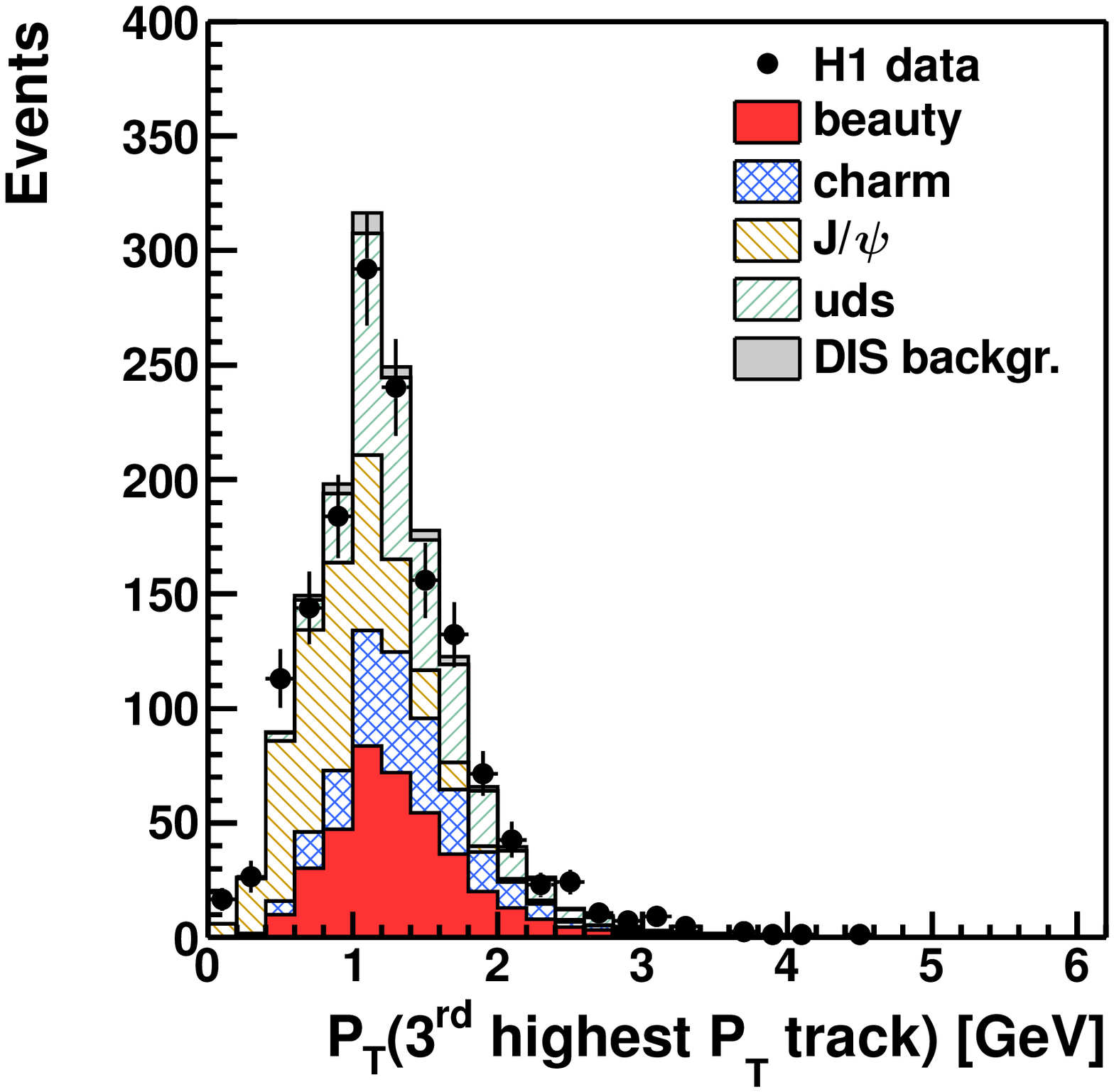,width=7.5cm}\put(-0.8,0.9) {c)}
\setlength{\unitlength}{1cm}
\caption{Control distributions for the three highest $P_T$-tracks of the hadronic final state  as
  function of the track $P_T$. 
Data are compared to the Monte Carlo simulations using the
the quark flavour decomposition determined by the unfolding procedure.
Data are represented as points with the
statistical uncertainties indicated by the error bars.
}
\label{fig:ControlDistrTracks} 
\end{figure}
\begin{figure}[htp]
\center
\epsfig{file=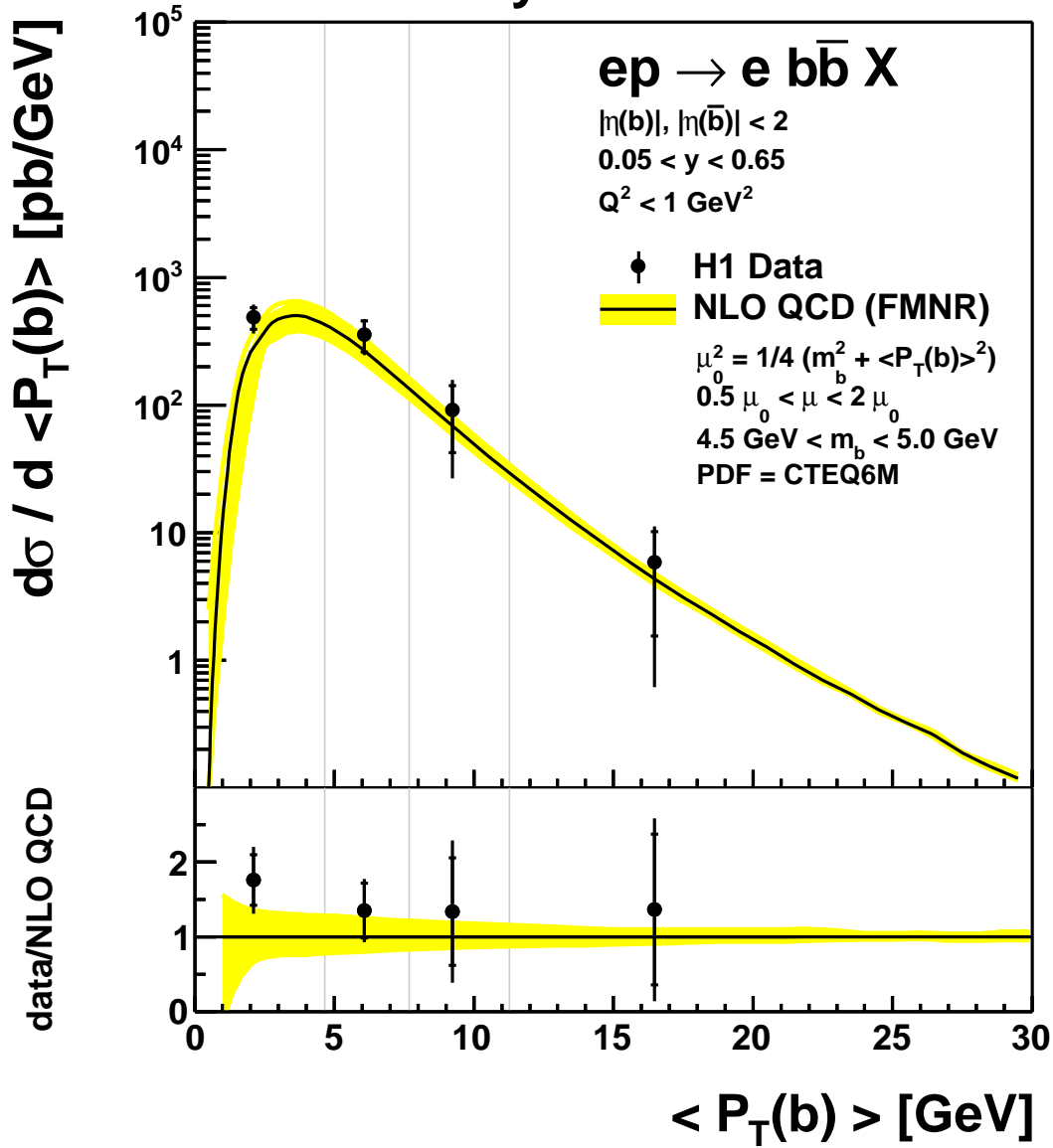 ,width=\textwidth}
\setlength{\unitlength}{1cm}
\caption{Differential beauty cross section \mbox{$\mathrm{d}\sigma / \mathrm{d} \langle P_T(b) \rangle$} 
 shown as function of the quadratically  averaged transverse momentum of the beauty quarks $\langle P_T(b) \rangle$
 (upper part). 
The data are represented by points with  inner vertical error bars
representing the statistical errors and  outer error bars representing the
total error. 
The vertical gray lines indicate the bin boundaries in $\langle P_T(b) \rangle$ of each
data point and the points are shown at the bin centred positions.
The data are compared to the FMNR NLO QCD calculation (solid line) with the
uncertainty represented as shaded band.
Also shown is the ratio of the measured cross section to the calculated
NLO QCD prediction, 
$ \frac{\mathrm{d}\sigma_{\mathrm{measured}}} {\mathrm{d} \langle P_T(b) \rangle} / \frac{\mathrm{d}\sigma_{\mathrm{NLO \ QCD}}}{\mathrm{d} \langle P_T(b) \rangle}$
(lower part).}
\label{fig:DiffXsectionPtB} 
\end{figure}

\end{document}